\documentclass[unnumsec,webpdf,modern,medium]{template}

\onecolumn
\usepackage{fancyhdr,amsmath,amssymb, amsthm,mathtools}
\usepackage{bbm,bm}
\usepackage{dsfont}
\usepackage{graphicx}
\usepackage{algorithm}
\usepackage[export]{adjustbox}
\usepackage[british]{babel}
\usepackage{calc} 
\setcitestyle{round}
\usepackage{adjustbox}
\usepackage{subcaption}
\usepackage{booktabs}
\usepackage{hyphenat}

\makeatletter
\patchcmd{\@@address}
  {\hbox}
  {\relax}
  {}{}
\renewcommand{\addressandsep}{\par}
\makeatother

\newlength{\leftcolumnwidth}
\newcounter{globalrem}

\newcounter{globaldef}

\newcounter{globalprop}

\newcounter{globalem}

\newcounter{globatheo}

\newcounter{globalcor}

\newcounter{globalas}

\begin{document}
\journaltitle{}
\copyrightyear{2022}
\pubyear{2019}
\appnotes{$_.$}

\firstpage{1}
\setcounter{secnumdepth}{2}

\fontsize{9.92}{10.48}\selectfont

\title[Bayesian local clustering of age-period mortality surfaces across multiple countries]{Bayesian local clustering of age-period mortality surfaces across multiple countries}

\author[1]{Giovanni Roman\`o}
\author[2]{Emanuele Aliverti}
\author[1,3]{Daniele Durante}

\authormark{Roman\`o, Aliverti and Durante}

\address[1]{\orgdiv{Department of Decision Sciences},  \orgname{Bocconi University},  \country{Via Roentgen 1, Milan, Italy}}
\address[2]{\orgdiv{Department of Statistical Sciences},  \orgname{University of Padova},  \country{Via Cesare Battisti 241, Padova, Italy}}
\address[3]{\orgdiv{Bocconi Institute for Data Science and Analytics},  \orgname{Bocconi University},  \country{Via Roentgen 1, Milan, Italy}}

\corresp[]{Address for correspondence: Giovanni Roman\`o, Department of Decision Sciences, Bocconi University, Via Roentgen 1, Milan, Italy. Email: \href{email: giovanni.romano4@phd.unibocconi.it}{giovanni.romano4@phd.unibocconi.it}}

\abstract{
\fontsize{9.6}{10}\selectfont
Although traditional literature on mortality modeling has focused on single countries in isolation, recent contributions have progressively moved toward joint statistical models for multiple countries. Besides favoring borrowing of information to improve age-period forecasts, this perspective has also potential to infer local similarities among countries’ mortality patterns within specific age classes and periods that could unveil unexplored demographic trends, while guiding the design of targeted policies. Advancements along this direction are currently undermined by the lack of a multi-country model capable of incorporating the core structures of age-period mortality surfaces together with clustering patterns among countries that are not global, but rather vary locally across the different combinations of age classes and periods. We cover this gap by developing a novel Bayesian model for log-mortality rates that characterizes the age structure of mortality through a B-spline expansion whose country-specific dynamic coefficients encode both changes of this age structure across periods and also local clustering patterns among countries under a time-dependent random partition prior for these country-specific dynamic coefficients. While flexible, this formulation facilitates tractable posterior inference leveraging a suitably designed Gibbs sampler. The application to mortality data from 14 countries unveils local similarities which highlight both previously recognized demographic phenomena and also yet unexplored patterns.
}


\maketitle
\fontsize{9.92}{10.47}\selectfont
\vspace{-10pt}

\section{\large 1. Introduction} \label{sec_1}
The changes in life expectancy, population structure and welfare systems over the past decades have stimulated a growing demand for new statistical models capable of characterizing heterogeneous mortality patterns across ages, periods and countries, while providing reliable probabilistic forecasts with rigorous uncertainty quantification \citep[see, e.g.][]{hunt2021structure,lutz2010,raftery2013}. Advancements along these lines are fundamental in guiding healthcare, social, environmental and retirement policies, thereby motivating an active research on mortality modeling in several fields, such as demography \citep[see, e.g.][]{Camarda2019, deJong2006, hyndman2013coherent, lee2001evaluating, Leger&Mazzucco2021, li2013extending, Li&Lee2005, mazzuco2018, raftery2013}, statistics \citep[see, e.g.][]{alexopoulos2019bayesian, Aliverti2022, debon2023multipopulation, dimai2025clustering, hyndman2007, lam2023multipopulation, Lee&Carter1992, Pavone&Legramanti&Durante2022, tang2022clustering} and actuarial sciences \citep[see, e.g.][]{antonio2015bayesian, currie2016fitting, dong2020multi, enchev2017multi, haberman2011comparative, hatzopoulos2013common, kleinow2015common, scognamiglio2022multi, wong2018bayesian}, among others. As highlighted in these contributions, such an active research on mortality has witnessed in recent years a progressive shift  away from analyzing single countries in isolation and towards joint modeling of age-period mortality surfaces within a multi-country setting. 

\begin{figure}[t]
\centering
   \centering
    \includegraphics[width =0.8\linewidth, center]{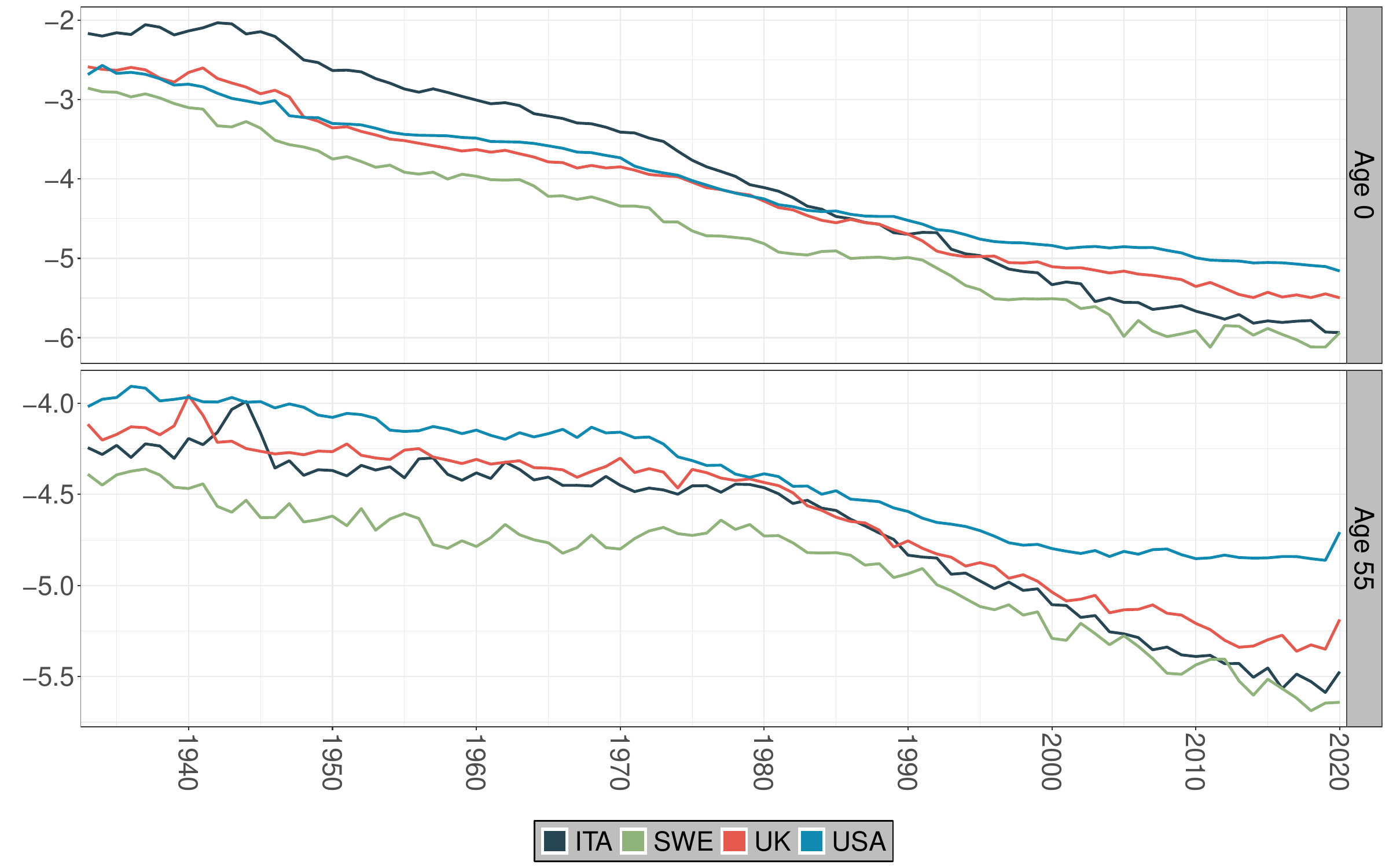}
    \caption{\footnotesize Log-mortality rates at age $0$ (first panel) and $55$ (second panel) in Italy (ITA), Sweden (SWE), the United Kingdom (UK) and the United States (USA) between 1933 and 2020.}
    \label{fig:0and55yrs_old_data}
\end{figure}

The above perspective is inherently motivated by the fact that the countries are not isolated entities, but rather display similar global patterns in age-specific mortality curves (see Figure~\ref{fig:0and55yrs_old_data}) regulated by common exogenous and endogenous factors, such as advancements in healthcare \citep[e.g.][]{vallin2004}. As such, joint modeling of multiple countries can facilitate effective borrowing of information to improve age-period mortality forecasts, while  opening the avenues for inference on local similarities and differences among countries’ mortality patterns in specific age classes and periods that could unveil unexplored demographic trends,  possibly arising from targeted policies adopted by certain countries. Addressing both objectives would require a unique formulation that accounts not only for the core structures of country-specific age-period mortality surfaces, but also for local clustering patterns between countries arising from overlaps among these country-specific surfaces in particular combinations of ages and periods. Although state-of-the-art multi-country mortality models are still not designed to include these fundamental dynamics, local heterogeneities in country-specific mortality rates across ages and periods are recognized in demographic research \citep[see, e.g.][]{vallin2004,vaupel2011} and find strong evidence within mortality data. For example, the first panel of Figure~\ref{fig:0and55yrs_old_data} illustrates the infant log-mortality rates within Italy, Sweden, the United Kingdom (UK) and the United States (USA) from $1933$ to $2020$, as retrieved from the Human Mortality Database (HMD: \url{https://www.mortality.org/}).  During the first fifty years, the UK and the USA display overlapping mortality trajectories, which then diverge after 1980. Conversely, Italy follows a distinct path until 1985, after which it aligns with the UK for approximately ten years.  Interestingly, as shown in the second panel of Figure~\ref{fig:0and55yrs_old_data}, these grouping structures vary not only across periods, but also with different ages. For instance, the clustering among the UK and the USA at infant ages is no longer visible for individuals aged 55, who show, instead, a clear overlap between Italy and the UK for a large time window.

Accounting for the above patterns via a principled model-based representation would not only provide a more realistic characterization of age-period mortality surfaces across multiple countries,  but could also open the avenues for rigorously answering important scientific questions, e.g., on the differences among low mortality countries \citep[][]{oeppen2002,vaupel2011} and the corresponding socio-economic determinants \citep{marmot2005}. As discussed previously, although the literature on multi-country mortality modeling has witnessed extensive advancements over the recent years, state-of-the-art contributions are not designed to address this endeavor. In fact, a common solution in multi-country mortality modeling relies on extending  bilinear  decompositions of age-period mortality surfaces for a single country, such as the one by \citet{Lee&Carter1992}, to joint formulations that include country-specific parameters and a common component shared across periods \citep[][]{Li&Lee2005} or ages \citep[][]{kleinow2015common}. This perspective improves forecasts and can be further extended to more sophisticated representations leveraging multi-country generalizations  \citep{hyndman2013coherent,lam2023multipopulation} of functional principal components constructions \citep{hyndman2007}, or  joint decompositions of the country-age-period mortality tensor \citep{dong2020multi}; see also \citet{enchev2017multi}. However, all these solutions are not designed to infer group structures among countries based on local similarities in the associated mortality rates. Such an issue is also found in related Bayesian hierarchical formulations, which rely on conditionally independent models for every country, linked by a common prior on shared  parameters \citep[e.g.][]{Aliverti2022, antonio2015bayesian, raftery2013}.

An effective direction for overcoming the aforementioned issues is to move towards more recent extensions of the above formulations which explicitly include notions of clustering among countries in terms of the corresponding mortality patterns \citep[e.g.][]{debon2023multipopulation, dimai2025clustering, hatzopoulos2013common, Leger&Mazzucco2021, schnurch2021clustering, scognamiglio2022multi, tang2022clustering}. Albeit providing meaningful representations for model-based clustering  of mortality surfaces, the overarching focus of these extensions is on global grouping structures, instead of local ones changing with different combinations of ages and periods.  As such, countries are not allowed to display varying clustering behaviors at different ages \citep{Leger&Mazzucco2021}, time periods \citep{hatzopoulos2013common}, or both dimensions \citep{dimai2025clustering, schnurch2021clustering, scognamiglio2022multi, tang2022clustering}. Recalling Figure~\ref{fig:0and55yrs_old_data}, these constraints are not supported by the observed age-period mortality  data and hinder the possibility of learning nuanced grouping structures, possibly informing on unexplored localized demographic patterns that are visible only for specific combinations of ages and periods.  Note also that these grouping structures should be expressed in terms of similarities with respect to both the overall level and trend in country-specific mortality rates. This is crucial in comparative mortality analysis, since populations may share similar temporal trends while remaining markedly different in terms of absolute mortality levels \citep[e.g.][]{debon2023multipopulation,Leger&Mazzucco2021}. Therefore, clustering  only on trend would group countries with parallel declines but persistently different mortality intensities, thereby failing to identify structurally different populations  requiring substantially different policy interventions.

To our knowledge, the only attempt to move beyond the aforementioned representations  can be found in the application of latent class clustering models to multi-country age-period  mortality data in \citet{debon2023multipopulation}. Although this contribution has the merit of recognizing the importance of moving beyond global clustering perspectives, a direct application of latent class clustering without the  inclusion of the specific structures of age-period mortality surfaces could undermine the flexibility of the resulting procedure in uncovering local clustering patterns. This issue finds evidence in Figure 1 of  \citet{debon2023multipopulation} where the inferred groups resemble more closely those obtained under a global perspective. For example, according to Figure~1 of  \citet{debon2023multipopulation}, at infant ages, all countries share the same cluster in the entire time window analyzed, a  behavior which is not in line with the local clustering structures displayed by the observed mortality data in our Figure~\ref{fig:0and55yrs_old_data}. Moreover, similarly to  cluster-based extensions of multi-country mortality models, also  \citet{debon2023multipopulation} requires specifying the unknown number of groups, a challenging task in practice, without a uniquely accepted solution.

Motivated by the above discussion and by the impact of addressing the aforementioned challenges in multi-country mortality modeling, we propose and develop in Section~\ref{section:model_formulation} an innovative and principled Bayesian formulation that accounts for the core structures of age-period mortality surfaces, and crucially incorporates clustering patterns among countries which are allowed to vary flexibly across  different combinations of ages and periods. This is accomplished by modeling the smooth age patterns of mortality through a flexible linear combination of B-spline bases \citep[][]{Camarda2019,Pavone&Legramanti&Durante2022}, whose country-specific dynamic coefficients evolve across periods via carefully designed stochastic processes of time relying on a temporal random partition prior inspired by the general construction of \citet{Page2022}. These stochastic processes for the joint time trajectories of the country-specific coefficients associated to the different B-spline bases provide a principled characterization of the time changes in the age patterns of mortality across periods, while allowing the grouping structures exhibited by countries to change both in time and across the basis coefficients associated with the different ages.  As such, the resulting representation facilitates the identification of locally converging or diverging patterns within multi-country age-period mortality surfaces, while preserving a structured representation that accounts for the core characteristics of these surfaces. 

As clarified in Section~\ref{sec_3}, the proposed representation, albeit flexible, is amenable to tractable posterior inference via a carefully designed Gibbs-sampling algorithm that learns automatically the unknown total number of clusters and facilitates both point estimation and uncertainty quantification on mortality patterns and grouping structures. The simulation studies in Section~\ref{sec_4} and the application to mortality data of 14 countries from 1933 until 2020 in Section~\ref{sec_5}, not only confirm the ability of the proposed model to accurately learn these local grouping structures, but also unveil localized similarities among specific countries, which highlight both known demographic phenomena and also unexplored patterns acting on given countries over specific combinations of ages and periods. Concluding remarks can be found in Section~\ref{sec_6}, where we also clarify that, although motivated by multi-country age-period mortality data, our model has broader scope, in that it allows to detect localized overlaps among generic surfaces from different populations.

\section{\large 2. Modeling age-period mortality surfaces across multiple countries} 
\label{section:model_formulation}
\subsection{\large 2.1 Model formulation} \label{subsection:model_formulation}
Let us denote with $d_{ixt}$ and $\textsc{e}_{ixt}$, respectively,  the observed death counts and the average number of individuals at risk within country $i = 1, \dots, n$, at age $x \in \mathcal{X}$ and for the period  $t=1,\dots,T$ (corresponding, in our case, to calendar years). Consistent with the overarching focus in mortality models for both single and multiple countries \citep[see, e.g.][]{currie2016fitting,enchev2017multi,hunt2021structure}, our interest lies in the analysis of  the country-specific age-period log-mortality rates $\log m_{ixt} = \log({d_{ixt}}/{\textsc{e}_{ixt}})$ for which we assume 
\begin{equation}
\log m_{ixt} = f_{it}(x)+\varepsilon_{ixt}, \qquad \mbox{with} \ \ \varepsilon_{ixt} \sim \mbox{N}(0, \sigma_{i}^2),
\label{eq_1}
\end{equation}
independently  for every $i = 1, \dots, n$, $x \in \mathcal{X}$ and $t=1,\dots,T$, where $f_{it}(x) = \mathbb{E}[\log m_{ixt} \mid f_{it}(x)]$ denotes the expected log-mortality rate surface for the $i$-th country expressed as a function of age $x \in \mathcal{X}$ that is allowed to vary across periods $t=1, \ldots, T$. The general surface plus Gaussian noise formulation within \eqref{eq_1} is common to several mortality models for both single and multiple countries \citep[e.g.][]{currie2016fitting,enchev2017multi,hunt2021structure}. In addition, as recently proved by \citet[][Proposition 2.1]{Pavone&Legramanti&Durante2022} under a single-country perspective, when $\textsc{e}_{ixt}$ is sufficiently large (as in standard demographic settings), the above formulation arises directly from an underlying Poisson log-normal model for the observed death counts $d_{ixt}$, which accounts for possible overdispersion; see also \citet{wong2018bayesian}. However, as discussed in Section~\ref{sec_1}, none of the available multi-country models provides a structured representation for $f_{it}(x)$ that incorporates the core age-period patterns of mortality, while allowing countries to cluster differently as both $x$ and $t$ vary. 

Addressing this gap would require the design of a structured representation for the expected log-mortality rate surfaces which ensures that $f_{it}(x)$ varies {\em (i)} smoothly as a function of age $x$, for every $i = 1, \dots, n$ and $t=1,\dots,T$, and {\em (ii)} dynamically with periods $t$, for each $i = 1, \dots, n$ and $x \in \mathcal{X}$, while  {\em (iii)} allowing  $f_{it}(x)$ and $f_{i't}(x)$ for any two generic countries $i$ and $i'$ to cluster (i.e. display similar values) only for those combinations of $x$ and $t$ in which there is an empirical evidence of overlapping patterns in the observed log-mortality rates of $i$ and $i'$. To this end, a natural option for addressing objective {\em (i)} in a way that facilitates also inclusion of {\em (ii)} and {\em (iii)} is to represent $f_{it}(x)$ through the B-spline expansion
\begin{equation}
 f_{it}(x) =\sum_{j=1}^p \beta_{ijt} g_j(x), \qquad  i = 1, \dots, n, \ x \in \mathcal{X}, \ t=1,\dots,T,
 \label{eq:f_it}
\end{equation}
where $g_1(x), \dots, g_p(x)$ denotes a set of common B-spline bases \citep[e.g.][]{eilers1996flexible} with associated coefficients $\beta_{i1t}, \ldots, \beta_{ipt}$ that are allowed to vary with countries $i=1, \ldots, n$ and periods $t=1, \ldots, T$. Although related representations have been traditionally explored in separate  analyses   of single countries in isolation \citep[see, e.g.][]{Camarda2019,currie2004,Pavone&Legramanti&Durante2022}, the expansion in \eqref{eq:f_it} provides an effective construction in our multi-country setting which ensures that $f_{it}(x)$ is a smooth function of age, for every $i = 1, \dots, n$ and $t=1,\dots,T$, whose dynamic changes across periods and local clustering patterns among countries can be regulated by a finite set of coefficients $\beta_{i1t}, \ldots, \beta_{ipt}$. Hence, objective  {\em (i)} is addressed by construction, while {\em (ii)}--{\em (iii)} can be accomplished by allowing time changes and ties in these coefficients, respectively.  More specifically, notice that, when $\beta_{ijt} = \beta_{i' jt}$, then countries $i$ and $i'$ are expected to display under   \eqref{eq:f_it} similar mortality rates within period $t$ for the age interval associated with  basis $g_j(x)$. This specification allows to disentangle model flexibility from its ability to characterize the heterogeneity across countries. The former is achieved through a careful choice of the B-spline bases to represent each country- and period-specific log-mortality curve $f_{it}(x)$ as a flexible function of age. Conversely, the latter is governed by the clustering on the basis coefficients. Few clusters induce homogeneous structures with most countries displaying similar mortality patterns, whereas a larger number of clusters leads to more heterogeneous structures differing across countries.

\begin{figure}[t]
    \centering
    \includegraphics[width = 0.95 \textwidth]{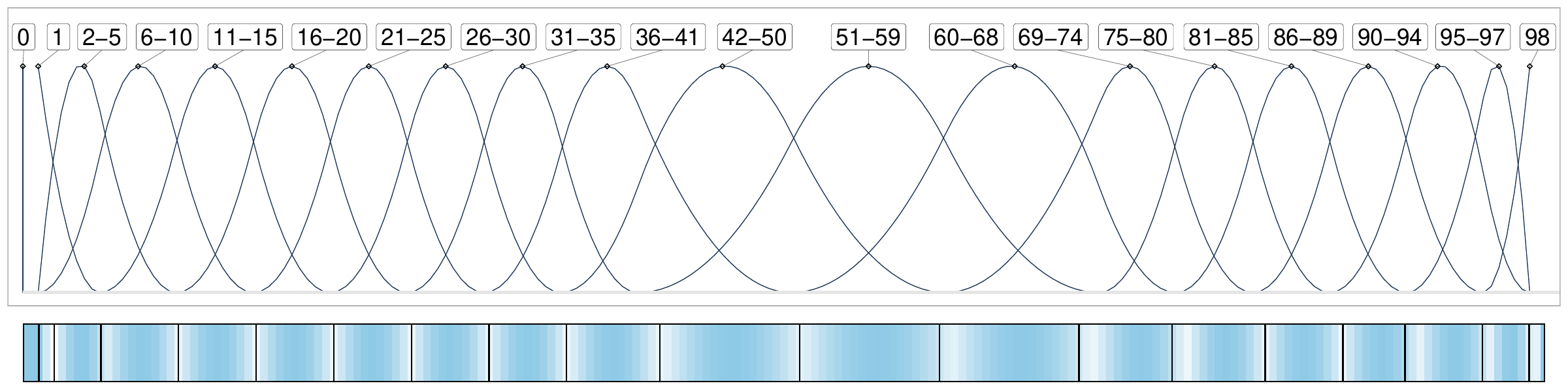}
    \caption{\footnotesize Graphical representation of the selected B-spline bases $g_1(x), \ldots, g_{20}(x)$. The numbers associated to each basis $g_j(x)$ denote the {age interval} at which such basis is {\em leading} (i.e. it dominates the others). The rectangles in the bottom part of the figure highlight the age intervals associated to the different spline bases. Colored areas represent the value of the {\em leading} basis in its corresponding age interval, with darker colors denoting higher values.}
    \label{fig:splines}
\end{figure}

Motivated by the above discussion, we focus on quadratic B-spline bases defined as in Figure \ref{fig:splines}. In recent single-country analyses \citep[][]{Pavone&Legramanti&Durante2022}, such a choice proved effective to achieve a realistic characterization of the age patterns of mortality, allowing increased flexibility in those age ranges where larger local variations are expected. In addition, the age intervals in which each basis dominates the others are disjoint and span roughly five years for most bases. This aligns with the resolution of mortality analyses in demographic modeling  \citep[e.g.][]{hmd,vsevvcikova2016age}, while achieving a more realistic smooth characterization via B-splines, instead of modeling  separately mortality  patterns over a fixed discretization into five-year classes. Furthermore, as discussed previously, it facilitates the inclusion of local clustering behaviors, both in the overall level and trend of  $f_{it}(x)$, since the intercept is implicit in its expansion. To formalize the cluster property,  let $c_{ijt} \in \{1, \ldots, K_{jt}\}$ be the group membership indicator for country $i$, with respect to the $j$-th  basis in period $t$. Moreover, for each $j=1, \ldots, p$ and $t=1, \ldots, T$, define with $\mathbf{c}_{jt} = [c_{1jt}, \dots, c_{njt}] \in \{1, \ldots, K_{jt}\}^{n}$ the vector comprising the memberships of the $n$ countries, where $K_{jt} \leq n$ is the total number of clusters at the pair $(j,t)$. Then, under~\eqref{eq:f_it}, local overlaps in the mortality curves can be induced through ties among the coefficients $\beta_{1jt}, \ldots, \beta_{njt}$, for every $j=1, \ldots, p$ and $t=1, \ldots, T$, by letting
\begin{equation}
 \beta_{ijt}=\beta^\star_{c_{ijt}jt}, \qquad \mbox{for all } i=1, \ldots, n, \ j=1, \ldots, p, \ t=1, \ldots, T,
 \label{eq:f_clust_beta_p}
\end{equation}
where $\beta^\star_{c_{ijt}jt} \in \mathbb{R}$ is the value of the $j$-th  B-spline coefficient in period $t$ associated with the cluster to which country $i$ has been allocated. As such, all countries belonging to the same generic cluster $k$ (for the $j$-th  basis in period $t$) will display the same value $\beta^\star_{kjt}$ for the associated coefficient, thereby addressing {\em (iii)}. Crucially, $\mathbf{c}_{jt}$ varies with both $j$ and $t$, and therefore any generic pair of countries  $i$ and $i'$ is allowed to cluster locally only for specific ages and periods, consistent with the original motivations underlying the proposed formulation.

In interpreting the form of clustering induced by the above construction, notice that, as shown in Figure~\ref{fig:splines}, due to the local support property of the selected B-spline, for any age $x \in \mathcal{X}$ three adjacent bases are active \citep[e.g.][]{eilers1996flexible}. Therefore,  equality between two country-specific mortality curves $f_{it}(x)$ and $f_{i^\prime t}(x)$ at age $x$ --- for a fixed calendar year $t$ --- is achieved when all the three coefficients associated with the active basis functions at that age $x$ coincide. Although this property may suggest caution in interpreting the clusters at the level of each spline coefficient as an exact overlap of the mortality curves in the age range where the basis is active, it shall be emphasized that the age support can be divided into non-overlapping intervals, each having a single basis that dominates the others. More formally,  for each $x \in \mathcal{X}$, it is possible to define the {\em leading} basis by the index $j^\star(x)= \min\{\mbox{argmax}_{j} g_{j}(x)\}$, and identify such a basis with the age interval $\mathcal{I}_j = [a_j, b_j)$ having limits $a_j = \mbox{inf}\{x: j^\star(x) = j\}$ and $b_j = \mbox{sup}\{x: j^\star(x) = j\}$; see Figure \ref{fig:splines} for a graphical illustration of the resulting age partition for the  B-spline bases employed in the proposed model. Under this perspective, clustering of the basis coefficients $\beta_{ijt}$ and $\beta_{i^\prime jt}$ is expected to induce similarities in the mortality curves $f_{it}(x)$ and $f_{i^\prime t}(x)$ over the age interval $\mathcal{I}_j$ in which the corresponding basis $g_j(x)$ is {\em leading}, thereby providing an effective approach for identifying common patterns. This intuition is reasonable in mortality data, since the underlying functions are smooth across both ages and periods. As  a result, abrupt changes in adjacent basis coefficients --- large enough to produce substantially different mortality curves even if the coefficients of the {\em leading} spline remain clustered --- are unlikely under both the proposed model and the demographic application considered. These intuitions are also confirmed empirically through careful quantitative assessments in Sections~\ref{sec_4}--\ref{sec_5}, supporting the effectiveness of the proposed strategy for interpreting the clustering results in mortality data.

\subsection{\large 2.2 Prior specification} \label{subsection:model_prior_spec}

In order to complete the proposed Bayesian formulation, we require priors for the parameters of the  model in \eqref{eq_1}--\eqref{eq:f_clust_beta_p}, namely $\sigma_i^2$, for each $i=1, \ldots, n$,  along with $\mathbf{c}_{jt} $ and $\boldsymbol{\beta}_{jt}^\star = [{\beta}_{1jt}^\star, \dots, {\beta}_{K_{jt}jt}^\star]^\intercal$, for every $j=1, \ldots, p$, $t=1, \ldots, T$. Regarding  the country-specific variances, we follow common practice and consider conditionally-conjugate $\mbox{Inv-Gamma}(a_\sigma, b_\sigma) $ priors for every $\sigma_i^2$, independently across $i=1, \ldots, n$. Conversely, for $\mathbf{c}_{jt} $ and $\boldsymbol{\beta}_{jt}^\star$ we elicit priors which explicitly account for the temporal structure behind the evolution of both  $\mathbf{c}_{jt} $ and $\boldsymbol{\beta}_{jt}^\star$ across periods $t=1, \ldots, T$, for each $j=1, \ldots, p$, thereby addressing {\em (ii)}. Focusing first on $\boldsymbol{\beta}_{jt}^\star$, this is accomplished by assuming independent Gaussian priors for the entries in such a vector, further centered around a higher-level mean function of time that is assigned a Gaussian process (GP) prior \citep[see, e.g.][]{williams2006gaussian}. More specifically, we let
\begin{equation}
\begin{split}
&(\boldsymbol{\beta}^\star_{jt} \mid \mathbf{c}_{jt}, \psi_{jt}, \delta_j^2)  \sim \mbox{N}_{K_{jt}}(\psi_{jt}\mathds{1}_{K_{jt}}, \delta_j^2 \mathbf{I}_{K_{jt}}), \quad \mbox{independently for } j = 1, \dots, p, \ t = 1, \ldots, T,\\
&      (\boldsymbol{\psi}_j=[\psi_{j1}, \ldots, \psi_{jT}]^{\intercal}  \mid \omega_j^2) \sim \mbox{N}_{T}(\boldsymbol{\mu}_j, \omega_j^2 \boldsymbol{\Sigma}),  \quad \mbox{independently for }  j = 1, \ldots, p,
        \label{eq:model_beta}
        \end{split}
\end{equation}
where $\mathds{1}_{K_{jt}}$ and $\mathbf{I}_{K_{jt}}$ denote the $K_{jt} \times 1$ vector of ones and the $K_{jt} \times K_{jt}$ identity matrix, respectively, whereas $\boldsymbol{\mu}_j$ and $\omega_j^2 \boldsymbol{\Sigma}$ are the mean vector and the covariance matrix induced by the GP prior on the finite time grid $t=1, \ldots, T$. For  $ \boldsymbol{\Sigma}$ we consider, in particular, a squared exponential correlation function \citep[see, e.g.][Section 4]{williams2006gaussian} which allows the dependence between the generic $\psi_{jt}$ and $\psi_{jt'}$ to progressively decrease as the distance between the time indexes $t$ and $t'$ increases. The mean vector $\boldsymbol{\mu}_j$ is instead elicited under a data-driven perspective (see Section~\ref{sec_4}), whereas  $\delta_j^2 $ and $\omega_j^2$ are assigned conditionally-conjugate $\mbox{Inv-Gamma}(a_\delta, b_\delta)$ and $ \mbox{Inv-Gamma}(a_\omega, b_\omega)$ hyperpriors, respectively, independently for every $j=1, \ldots, p$.  Notice that available demographic knowledge could, in principle, be incorporated through informative priors or via suitable structural constraints; for example, the mortality rates are generally expected to decline over periods. However, besides complicating model formulation and prior elicitation, these constraints might be overly restrictive when modeling several countries jointly over a large temporal window. In these contexts, events such as pandemics or wars could introduce shocks that substantially deviate from the standard age-period pattern of mortality curves, thereby introducing bias in the analysis. For these reasons, we avoid such restrictions by relying on flexible basis functions whose dynamic coefficients are endowed with priors  that allow to effectively learn from the observed data both standard age-period patterns of mortality surfaces and also, when necessary, deviations from these patterns.

As discussed previously, temporal dependence is expected also in the sequence of cluster assignment vectors $\mathbf{c}_{j1}, \ldots, \mathbf{c}_{jT} $, for $j=1, \ldots, p$, which control heterogeneity of mortality surfaces across countries. In particular, it is reasonable to assume that the generic $\mathbf{c}_{j,t-1} $  influences the formation of $\mathbf{c}_{jt} $ through a mechanism where only a subset of countries can change the corresponding cluster membership when moving from $t-1$ to $t$, whereas the others preserve it. This Markovian  structure  implies that $\mbox{pr}(\mathbf{c}_{j1}, \ldots, \mathbf{c}_{jT})=\mbox{pr}(\mathbf{c}_{j1})\mbox{pr}(\mathbf{c}_{j2}  \mid \mathbf{c}_{j1} ) \cdots \mbox{pr}(\mathbf{c}_{jT}  \mid \mathbf{c}_{j, T-1} ) $, and therefore the mechanism defining $\mbox{pr}(\mathbf{c}_{jt}  \mid \mathbf{c}_{j,t-1} )$ should be combined with a careful prior on the initial membership vector $\mathbf{c}_{j1}$ at time $t=1$, which induces a flexible characterization of the joint prior over the entire sequence $\mathbf{c}_{j1}, \ldots, \mathbf{c}_{jT} $, for each $j=1, \ldots, p$. A construction of this type can be found in the temporal random partition ($t$RPM) prior recently proposed by \citet{Page2022}. Adapting this general construction to our specific context, we let
\begin{equation}
([\mathbf{c}_{j1}, \dots, \mathbf{c}_{jT}] \mid \alpha_j, M_j) \sim t\mbox{RPM}(\alpha_j, M_j), \qquad \mbox{independently for } j = 1,\dots,p,
        \label{eq:model_t_rand}
\end{equation}
where $\alpha_j \in [0,1]$ corresponds to a temporal persistence parameter controlling the transition mechanism behind $\mbox{pr}(\mathbf{c}_{jt}  \mid \mathbf{c}_{j,t-1} )$, whereas $M_j \in \mathbb{R}_+$ regulates the formation of $\mathbf{c}_{j1}$ through $\mbox{pr}(\mathbf{c}_{j1})$, which in turn influences those of the subsequent vectors $\mathbf{c}_{j2}, \ldots, \mathbf{c}_{jT} $ under the Markovian structure discussed above. Specializing \citet{Page2022} to our setting, such a formation mechanism for $\mathbf{c}_{j1}$ is assumed to be driven by a Chinese restaurant process (CRP) with parameter $M_j$. This prior belongs to the general Gibbs-type class \citep[e.g.][]{de_2013} and provides a flexible characterization for the formation of the grouping structures in $\mathbf{c}_{j1}$ driven by a tractable urn scheme. In particular,  let  $\mathbf{c}_{j1(-i)}=[c_{1j1}, \ldots, c_{i-1,j1}, c_{i+1,j1}, \ldots, c_{nj1} ]$ be the cluster membership vector for the $j$-th spline basis at time $t=1$, excluding the generic $i$-th country, and denote with $n_{kj1(-i)}$ and $K_{j1(-i)}$ the cardinality of cluster $k$ and the total number of non-empty clusters in $\mathbf{c}_{j1(-i)}$, respectively. Then, under this scheme, the prior on the cluster memberships for country $i$, given those of the others,  coincides with 
\begin{equation}\label{eq:predictive_CRP}
    \mbox{pr}(c_{ijt} = k \mid {\bf c}_{jt(-i)}, M_j) \propto \begin{cases}
        n_{kjt(-i)} & \quad k = 1, \dots, K_{jt(-i)}, \\ 
        M_j &\quad k = K_{jt(-i)} + 1.
    \end{cases}
\end{equation} 
As such, the generic country $i$ can either be allocated to an already occupied cluster with conditional probability proportional to its cardinality, or create a new one with probability regulated by the CRP parameter $M_j$. Besides illustrating the tractability of this construction, along with the role of  $M_j$, the above urn scheme also clarifies that the total number of clusters in $\mathbf{c}_{j1}$ can be learned automatically, without the need to pre-specify it. This is a key advantage compared to state-of-the-art cluster-based models for mortality \citep[][]{debon2023multipopulation, dimai2025clustering, hatzopoulos2013common, Leger&Mazzucco2021, schnurch2021clustering, scognamiglio2022multi, tang2022clustering} that cannot learn the number of clusters as part of the estimation process.

Under \eqref{eq:model_t_rand}, the above prior for $\mathbf{c}_{j1}$ is combined with a similarly tractable assignment mechanism regulating the formation of $\mathbf{c}_{j2}, \mathbf{c}_{j3}, \ldots, \mathbf{c}_{jT}$ via the generic conditional law $\mbox{pr}(\mathbf{c}_{jt}  \mid \mathbf{c}_{j,t-1} )$. This is accomplished by sampling a binary latent indicator variable $\gamma_{ijt} \sim \mbox{Bern}(\alpha_j)$ which controls, for every country $i=1, \ldots, n$,  whether  that country belongs to the same cluster when moving from $ \mathbf{c}_{j,t-1}$ to  $\mathbf{c}_{jt}$ (i.e. $\gamma_{ijt}=1$), or can possibly change its current allocation (i.e. $\gamma_{ijt}=0$); by construction, for the first time point $\gamma_{ij1} = 0$ for each $i = 1, \dots, n$. Given $\boldsymbol{\gamma}_{jt}=[\gamma_{1jt}, \ldots, \gamma_{njt}]$ and $ \mathbf{c}_{j,t-1}$, the allocation vector $\mathbf{c}_{jt}$ at time $t$ is sampled from the set of allocations compatible with $ \mathbf{c}_{j,t-1}$, under $\boldsymbol{\gamma}_{jt}$, namely all those allocation vectors yielding a partition of the $n$ countries that can be obtained from the one encoded within $ \mathbf{c}_{j,t-1}$ by reallocating only those countries for which $\gamma_{ijt}=0$. Under this sampling mechanism, each allocation vector within the compatible set is assigned a probability proportional to the one defined by the prior at time $t=1$. Besides its simplicity, this construction provides an effective and flexible representation whose temporal persistence is directly regulated by $\alpha_j$ (the higher $\alpha_j$ is, the more similar $ \mathbf{c}_{j,t-1}$ and $ \mathbf{c}_{jt}$ are). Inheriting the theory by \citet{Page2022} within our specific setting, this construction further ensures that the marginals for $\mathbf{c}_{j2}, \mathbf{c}_{j3}, \ldots, \mathbf{c}_{jT}$ have the same CRP prior as the one assumed above for $\mathbf{c}_{j1}$, thereby yielding a temporal construction which, a priori, preserves for its marginals all the advantages of the CRP formulation discussed above for $ \mathbf{c}_{j1}$. 

The above property also guarantees that the temporal persistence of the process (regulated by $\alpha_j$) does not affect, a priori, the expected number of clusters in each time point (driven by $M_j$). This facilitates  both interpretation and learning  of the parameters $\alpha_j$ and $M_j$, $j=1, \ldots, p$, for which we consider  independent $\mathrm{Beta}(a_{\alpha}, b_{\alpha})$ and $\mathrm{Gamma}(a_{M}, b_{M})$ hyperpriors, respectively. The $\mathrm{Beta}(a_{\alpha}, b_{\alpha})$ hyperprior is employed also in \citet{Page2022}, while the $\mathrm{Gamma}(a_{M}, b_{M})$ for the CRP concentration parameters, although not present in  \citet{Page2022}, is motivated by recent theoretical results in  \citet{ascolani2023} on the consistency  properties of   CRP  constructions. Inference on such quantities allows the model to learn automatically, for each basis $j$, the optimal level of persistence of the clusters across periods and the concentration of the CRP based on the observed data. Section~\ref{sec_3} clarifies that the above Bayesian model is also amenable to tractable posterior computation and inference leveraging a carefully designed Gibbs-sampling algorithm.

\section{\large 3. Bayesian computation and inference}\label{sec_3}
Posterior inference for the Bayesian model in Section~\ref{section:model_formulation} is conducted via Monte Carlo leveraging the draws of a carefully designed Gibbs sampler targeting the  posterior distribution of the model parameters given the observed mortality rates. 
Section~\ref{sec:gibbs} derives in detail such a Gibbs sampler, while Section~\ref{sec:estimation} illustrates how  posterior samples are leveraged to perform Monte Carlo inference on the clustering structures among countries and the corresponding mortality patterns.

\subsection{\large 3.1 Gibbs sampler} \label{sec:gibbs}
The proposed Gibbs sampler iterates sequentially between two main steps. First, the temporal cluster memberships are sampled, along with the $t$RPM hyperparameters in \eqref{eq:model_t_rand}, from the corresponding full conditional distributions by adapting the algorithm of \citet[Section~B of the Supplementary Material]{Page2022} to our B-splines construction,  and the results in \citet{escobar1995} for the CRP hyperparameters $M_j$, $j=1, \ldots, p$. Second, conditionally on the group allocations, the cluster-specific B-spline coefficients in \eqref{eq:model_beta} and the corresponding hyperparameters are updated leveraging Gaussian and inverse-Gamma conjugacy.

In order to update the temporal grouping structures in $\mathbf{c}_{j1}, \ldots, \mathbf{c}_{jT} $, for each basis $j=1, \ldots, p$, along with the hyperparameters of the associated $t$RPM prior, let us first recall that the generic allocation vectors  $ \mathbf{c}_{j,t-1}$ and $ \mathbf{c}_{jt}$ are compatible, with respect to $\boldsymbol{\gamma}_{jt}$, if the partition of the $n$ countries encoded within $ \mathbf{c}_{jt}$ may be derived from the one associated with $ \mathbf{c}_{j,t-1}$ by reallocating only those countries for which $\gamma_{ijt}=0$. Given $ \mathbf{c}_{j,t-1}$ and $\boldsymbol{\gamma}_{jt}$, let us define with $\mathbb{C}(\mathbf{c}_{j,t-1}, \boldsymbol{\gamma}_{jt})$ the set comprising all partitions induced by $ \mathbf{c}_{jt}$ that are compatible with $\mathbf{c}_{j,t-1}$, under $\boldsymbol{\gamma}_{jt}$, and let $\Gamma_{jt} = \{i = 1, \dots, n : \gamma_{ijt} = 1\}$ be the set of countries whose group allocation does not change from $t-1$ to $t$. Furthermore,  denote with $\mathbbm{c}_{jt, \Gamma_{jt}}$ the partition induced by $ \mathbf{c}_{jt}$, but considering only those countries having indexes in $\Gamma_{jt}$. Then, leveraging the results in  \citet{Page2022}, the full-conditional distribution for the latent indicators $\gamma_{ijt}$  is a Bernoulli variable with probabilities 
\begin{equation}
    \mbox{pr}(\gamma_{ijt} = 1 \mid -) =\frac{\alpha_{j}\mathds{I}[\mathbbm{c}_{j, t-1,\Gamma_{jt(+i)}} = \mathbbm{c}_{j t,\Gamma_{jt(+i)}}]}{\alpha_{j} + (1 - \alpha_{j}) \mbox{pr}(\mathbbm{c}_{jt,\Gamma_{jt(+i)}}\mid M_j)/ \mbox{pr}(\mathbbm{c}_{jt,\Gamma_{jt(-i)}} \mid M_j)},
    \label{eq_gamma_latent}
\end{equation}
for every $i=1, \ldots, n$, $j=1, \ldots, p$ and $t=2, \ldots, T$, where $\mathds{I}[\cdot]$ denotes the indicator function, while $\Gamma_{jt(-i)} = \Gamma_{jt} \setminus \{i\}$ and $\Gamma_{jt(+i)} = \Gamma_{jt(-i)} \cup \{i\}$.  

Notice that in~\eqref{eq_gamma_latent}, the ratio $\mbox{pr}(\mathbbm{c}_{jt,\Gamma_{jt(+i)}} \mid M_j)/ \mbox{pr}(\mathbbm{c}_{jt,\Gamma_{jt(-i)}} \mid M_j)$ can be computed in closed-form leveraging the results in \citet{Page2022} under the urn scheme of the CRP prior discussed in Section \ref{section:model_formulation}. This is also a key to update the allocations $c_{ijt}$ of those countries for which $\gamma_{ijt}=0$; if the sampled $\gamma_{ijt}$ is $1$, then $c_{ijt}$ is kept fixed at the allocation drawn for $i$ at the pair $(j, t-1)$. To this end, let ${\bf c}_{jt(-i)}$ be the vector of cluster allocations  after removing the entry $c_{ijt}$, and denote with ${\bf c}_{jt(c_{ijt}=k)}$ the membership vector $[c_{1jt}, \ldots, c_{ijt}=k, \ldots, c_{njt}]$. Furthermore, let $K_{jt(-i)}$ be the total number of unique clusters in ${\bf c}_{jt(-i)}$ and define with $r_{ixt(j)} = \log m_{ixt} - \sum_{j' \neq j} \beta^\star_{c_{ij't}j' t} g_{j'}(x)$ the partial residuals under model \eqref{eq_1}--\eqref{eq:f_clust_beta_p}, without considering the $j$-th spline. Then,  leveraging the Bayes rule and the CRP urn scheme, we have that the full-conditional distribution for those $c_{ijt}$ with $\gamma_{ijt}=0$ is a categorical variable having probabilities, for every $k=1, \ldots, K_{jt(-i)}+1$, defined as
\begin{equation}\label{eq:fullcond_cluster}
\begin{split}
&\mbox{pr}(c_{ijt} = k \mid -) \propto \\
&  \quad \mbox{pr}(c_{ijt} = k \mid {\bf c}_{jt(-i)}, M_j) \mathds{I}[\mathbbm{c}_{j,t+1}\in \mathbb{C}({\bf c}_{jt(c_{ijt}=k)}, \boldsymbol{\gamma}_{j,t+1})]\prod_{x \in \mathcal{X}} \phi(r_{ixt(j)} - \beta^\star_{kjt}g_j(x); \sigma^2_i),
    \end{split}
\end{equation}
for each $i=1, \ldots,n$, $j=1, \ldots,p$ and $t=1, \ldots, T$, where $\phi(r_{ixt(j)} - \beta^\star_{kjt}g_j(x); \sigma^2_i)$ is the density, evaluated at $r_{ixt(j)} - \beta^\star_{kjt}g_j(x)$, of the zero-mean Gaussian distribution with variance $\sigma^2_i$, whereas, as discussed in Section~\ref{section:model_formulation}, $\mbox{pr}(c_{ijt} = k \mid {\bf c}_{jt(-i)}, M_j)$ is defined via the urn scheme in \eqref{eq:predictive_CRP}.

Given the samples of $\mathbf{c}_{jt}$ and $\boldsymbol{\gamma}_{jt}$, for $t=2, \ldots T$,  the $t$RPM  hyperparameters $\alpha_{j}$ are updated from the full-conditional Beta distributions 
\begin{equation}
    (\alpha_j\mid-) \sim \mathrm{Beta}\left(a_\alpha + \sum_{i = 1}^n \sum_{t = 2}^T \gamma_{ijt}, b_\alpha + n(T-1) - \sum_{i = 1}^n \sum_{t = 2}^T \gamma_{ijt}\right),
\end{equation}
independently for $j=1, \ldots, p$, whereas the CRP concentration parameters $M_j$, $j=1, \ldots, p$, are updated following the data-augmentation scheme described in \citet{escobar1995}.

Once the group membership vectors $\mathbf{c}_{j1}, \ldots, \mathbf{c}_{jT} $ have been updated for each $j=1, \ldots, p$, it is possible to sample the cluster-specific coefficients in $\boldsymbol{\beta}_{jt}^\star$, for each $j=1, \ldots, p$ and $t=1, \ldots, T$, along with the corresponding  means in   $\boldsymbol{\psi}_j$, for $j=1, \ldots, p$. Combining prior \eqref{eq:model_beta} with the model \eqref{eq_1}--\eqref{eq:f_clust_beta_p}, this can be accomplished by leveraging Gaussian-Gaussian conjugacy. In particular, let $\omega_{\beta_{kjt}^\star} = 1/\delta_j^2 + \sum_{x \in \mathcal{X} } \sum_{i : c_{ijt}=k}  (g_j^2(x)/\sigma_i^2)$ and $\eta_{\beta_{kjt}^\star} = (\psi_{jt}/\delta_j^2) +  \sum_{x \in \mathcal{X}} \sum_{i : c_{ijt}=k} r_{ixt(j)} g_j(x)/\sigma_i^2$, then the full-conditional distribution for the generic $\beta_{kjt}^\star$ is 
\begin{equation}
\label{full_beta_cond}
(\beta_{kjt}^\star\mid-) \sim \mbox{N}(\omega_{\beta_{kjt}^\star}^{-1}\eta_{\beta_{kjt}^\star},\omega_{\beta_{kjt}^\star}^{-1}),
\end{equation} 
independently for $k=1, \ldots, K_{jt}$, $j=1, \ldots, p$ and $t=1, \ldots, T$. Similarly, the full-conditional for each vector $\boldsymbol{\psi}_j$ is 
\begin{equation}
(\boldsymbol{\psi}_j\mid -) \sim \mbox{N}_T(\boldsymbol{\Omega}_{\psi_j}^{-1}\boldsymbol{\eta}_{\psi_j}, \boldsymbol{\Omega}_{\psi_j}^{-1}),
\label{full_psi_cond}
\end{equation}
for $j=1, \ldots, p$, where $\boldsymbol{\Omega}_{\psi_j} = \omega_j^{-2}\boldsymbol{\Sigma}^{-1} + \delta_j^{-2}\mbox{diag}(K_{j1}, \dots, K_{jT})$ and $\boldsymbol{\eta}_{\psi_j} = \omega_j^{-2}\boldsymbol{\Sigma}^{-1} \boldsymbol{\mu}_j +\delta_j^{-2}\bar{\boldsymbol{\beta}}_{j}$, where $\bar{\boldsymbol{\beta}}_{j}$ is a vector of dimension $T \times 1$ having generic $t$-th entry $\sum_{k} \beta_{kjt}^\star$.

To conclude the Gibbs sampler it remains to update the variance parameters $\sigma_i^2$, $i=1, \ldots, n$ in \eqref{eq_1}, along with $\delta_j^2$ and $\omega_j^2$, $j=1, \ldots, p$, entering the priors in  \eqref{eq:model_beta}. By conditioning on the quantities sampled in \eqref{eq:fullcond_cluster} and \eqref{full_beta_cond}--\eqref{full_psi_cond}, the updates for these variance parameters follow directly from inverse-Gamma conjugacy properties, thereby obtaining
\begin{equation}
    (\sigma_i^2 \mid -) \sim \mbox{Inv-Gamma}\left(a_\sigma +\frac{XT}{2}, b_\sigma + \frac{1}{2}\sum_{x \in \mathcal{X}} \sum_{t = 1}^T [\log m_{ixt} - f_{it}(x)]^2\right),
\end{equation}
for every $i=1, \ldots, n$, with $X$ denoting the cardinality of $\mathcal{X}$ and $f_{it}(x)$ defined as in \eqref{eq:f_it}--\eqref{eq:f_clust_beta_p}, and
\begin{equation}
\begin{split}
 &  (\delta_j^2 \mid -) \sim \mbox{Inv-Gamma}\left(a_\delta + \sum\nolimits_{t = 1}^T \frac{K_{jt}}{2}, b_\delta + \frac{1}{2}\sum_{t = 1}^T \sum_{k = 1}^{K_{jt}} (\beta_{kjt}^\star - \psi_{jt})^2\right), \\
&    (\omega_j^2 \mid -) \sim \mbox{Inv-Gamma}\left(a_\omega + \frac{T}{2}, b_\omega +  \frac{(\boldsymbol{\psi}_j - \boldsymbol{\mu}_j)^{\intercal} \boldsymbol{\Sigma}^{-1} (\boldsymbol{\psi}_j - \boldsymbol{\mu}_j)}{2}\right),
\end{split}
\end{equation}
for every $j=1, \ldots, p$.

\subsection{\large 3.2 Monte Carlo inference} \label{sec:estimation}
Leveraging the samples produced by the Gibbs routine outlined in Section~\ref{sec:gibbs}, posterior inference on the quantities of interest proceeds via Monte Carlo. As discussed in Sections~\ref{sec_1}--\ref{section:model_formulation}, within our context a primary focus is on inferring grouping structures among countries induced by similarities among the corresponding mortality rates, and how these group structures vary locally across ages and periods. This information is contained in the posterior samples of the allocation vectors $\mathbf{c}_{jt}$ for $j=1, \ldots, p$  and $t=1, \ldots, T$,  which we summarize through the $n \times n$ {\em posterior co-clustering matrices} $\hat{\bf P}_{jt}$, $j=1, \ldots, p$, $t=1, \ldots, T$, whose generic element $\hat{\bf P}_{jt[i,i']}$ is defined as the relative proportion of the Gibbs samples in which countries $i$ and $i'$ have the same group allocation, at the pair $(j,t)$. This provides an estimate of  posterior probabilities of co-clustering, which is not affected by label switching issues and is useful for quantifying uncertainty in $\mathbf{c}_{jt}$ for $j=1, \ldots, p$, $t=1, \ldots, T$, beyond single point estimates. As such, these matrices will be object of study in the mortality data application in Section~\ref{sec_5}.

When a single point estimate  $\hat{\mathbf{c}}_{jt}$ of $\mathbf{c}_{jt}$ is of interest for each $j=1, \ldots, p$ and $t=1, \ldots, T$, the above {\em posterior co-clustering matrices} can be summarized under the   decision-theoretic framework of \citet{Wade&Ghahramani2018} to obtain the estimate
\begin{equation*}
    \hat{\mathbf{c}}_{jt}= \mbox{argmin}_{{\mathbf{c}}'_{jt}} \mathbb{E}_{{\mathbf{c}}_{jt}}[\mbox{VI}({\mathbf{c}}_{jt}, {\mathbf{c}}'_{jt}) \mid \log \boldsymbol{m}],  \qquad j=1, \ldots, p, \ t=1, \ldots, T,
\end{equation*}
where $\log \boldsymbol{m}$ is the array of log-mortality rates observed for the $n$ countries across all ages and periods, whereas VI is the variation of information distance \citep{Meila2007}, namely a metric measuring the dissimilarity among two generic allocation vectors ${\mathbf{c}}_{jt}$ and ${\mathbf{c}}'_{jt}$ based on the associated individual and joint entropies. More specifically, dropping the indexes $j$ and $t$ for notational simplicity, the VI between  ${\mathbf{c}}$ and ${\mathbf{c}}^\prime$ with elements $c_i \in \{1, \ldots, K\}$ and $c^\prime_i \in \{1,\ldots,K^\prime\}$, is a function of the counts $n_{kk^\prime} = \sum_{i=1}^n \mathds{I}[c_i = k, c^\prime_i = k^\prime]$, $n_{k+} = \sum_{k^\prime=1}^{K^\prime}n_{kk^\prime}$,  $n_{+k^\prime} = \sum_{k=1}^{K}n_{kk^\prime}$ defined as 
\begin{equation*}
\mbox{VI}({\mathbf{c}}, {\mathbf{c}}') = -\sum_{k=1}^K \frac{n_{k+}}{n} \log_2\left(\frac{n_{k+}}{n}\right) -\sum_{k^\prime=1}^{K^\prime} \frac{n_{+k^\prime}}{n} \log_2\left(\frac{n_{+k^\prime}}{n}\right)-2\sum_{k=1}^{K}\sum_{k^\prime=1}^{K^\prime} \frac{n_{kk^\prime}}{n} \log_2\left(\frac{n_{kk^\prime} n}{n_{k+}n_{+k^\prime}}\right).
\end{equation*}
The above expression only depends on co-clustering relations, thereby circumventing  label switching issues. In practice, its minimization is solved, for each $j=1, \ldots, p$ and $t=1, \ldots, T$, via the \verb|R| package \verb|mcclust.ext| \citep[see,][]{Wade&Ghahramani2018} leveraging, as input, the corresponding {\em posterior co-clustering matrices} $\hat{\bf P}_{jt}$, $j=1, \ldots, p$, $t=1, \ldots, T$.

Besides the grouping structures encoded in $\mathbf{c}_{j1}, \ldots, \mathbf{c}_{jT} $, for $j=1, \ldots, p$, it is also of interest to study the trajectories $\beta_{ij1}, \ldots, \beta_{ijT}$ of the B-spline coefficients.  Recalling \eqref{eq:f_it}, these trajectories measure the temporal evolution of the mortality levels for country $i$ at the age interval associated with the $j$-th spline. The posterior samples for these quantities can be derived directly from those for $\mathbf{c}_{jt}$ and $\boldsymbol{\beta}_{jt}^\star$, after noticing that, under \eqref{eq:f_clust_beta_p}, $ {\beta}_{ijt}= {\beta}^\star_{c_{ijt}jt}$, for each $i=1, \ldots, n$, $j=1, \ldots, p$ and $t=1, \ldots, T$. As such, point estimates for the dynamic country-specific  B-spline coefficients can be derived via Monte Carlo by computing the average of the resulting samples for every $\beta_{ijt}$, $i=1, \ldots, n$, $j=1, \ldots, p$ and $t=1, \ldots, T$.

\section{\large 4. Simulation study}\label{sec_4}
Before employing the proposed Bayesian model in the motivating mortality data application, we  assess its performance in two simulation studies. The first scenario focuses on recovering the {\em true} data-generative structures  under correct specification. Recalling Sections~\ref{sec_1}--\ref{sec_3}, the overarching focus is on quantifying to what extent the proposed model can learn accurately realistic group structures that vary locally, along with the associated cluster-specific parameters, in different combinations of the ages and periods. The second scenario reproduces instead realistic mortality structures generated from a popular demographic model. In this setting, our proposed formulation is misspecified, and the interest lies in evaluating whether it can still provide an accurate fit to the simulated data, while achieving a sensible recovery of grouping structures, even if such structures do not arise, in the {\em true} data-generative mechanism, from clustering among spline coefficients.

\subsection{\large 4.1 First scenario}\label{sec_41}
\begin{figure}[t]
    \centering
    \includegraphics[width =0.95 \linewidth]{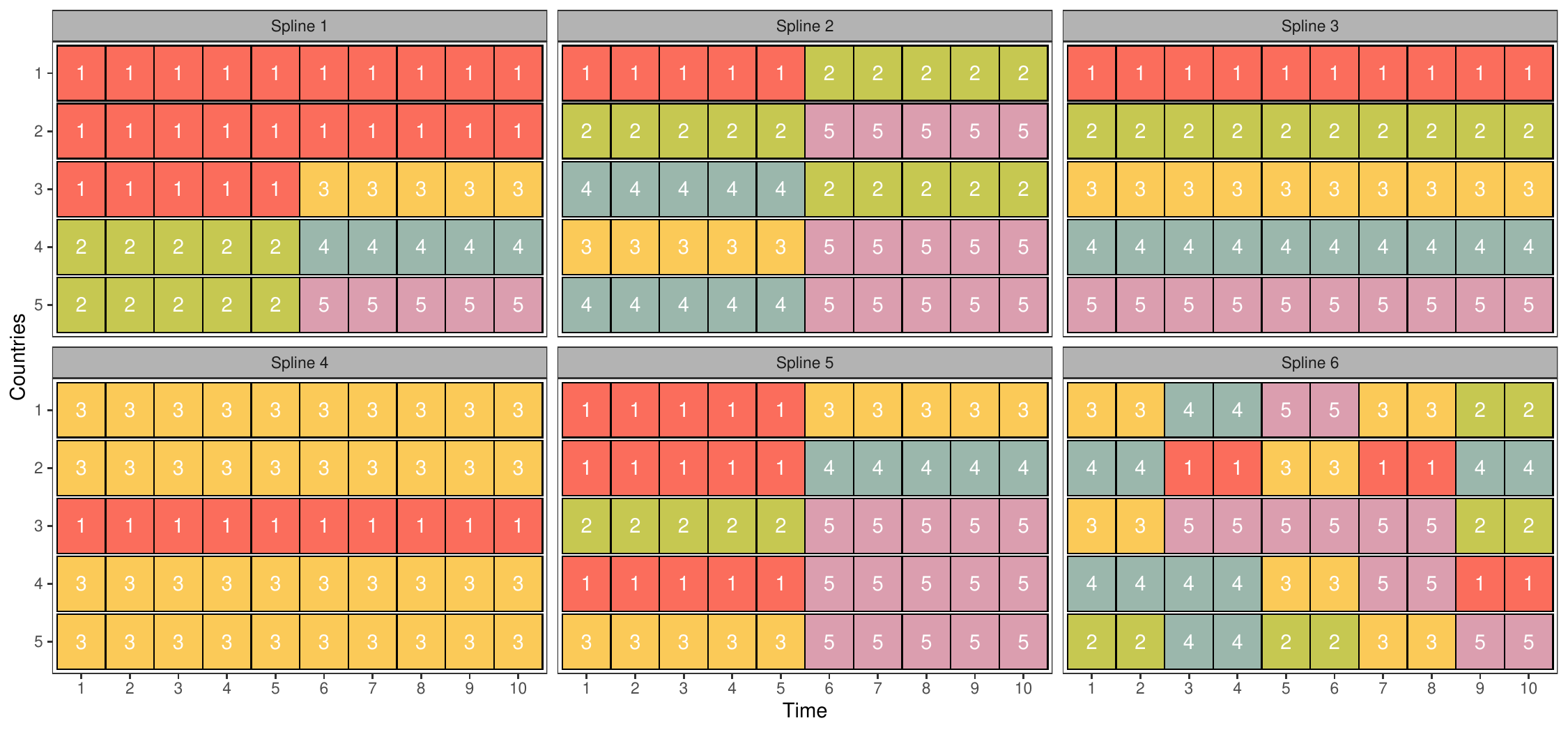}
    \caption{\footnotesize Cluster assignments in the simulation study. Colors and numbers represent true cluster memberships.}
    \label{fig:sim_study:clust_assign}
\end{figure}

We simulate synthetic log-mortality rates, under the model outlined in \eqref{eq_1}--\eqref{eq:f_clust_beta_p} focusing on $n = 5$ countries across $T = 10$ periods and for ages $\mathcal{X} =\{0, 1, \ldots, 100 \}$. More specifically, $\log m_{ixt}$ are simulated as in~\eqref{eq_1} with $\sigma_i = \sigma =  0.05$, for each $i=1, \ldots, 5$, and $f_{it}(x)$ defined through~\eqref{eq:f_it}--\eqref{eq:f_clust_beta_p} considering $p=6$ quadratic B-splines whose cluster-specific coefficients are generated under \eqref{eq:model_beta}, setting $\delta_j=0.05$ for $j\neq 5$, $\delta_5 = 0.1$, and letting $\boldsymbol{\psi}_j$, $j=1, \ldots, p$ correspond to parallel decreasing lines with slope $-0.02$. To assess the ability of the proposed model in inferring complex local group structures among countries, we do not produce $\mathbf{c}_{j1}, \ldots, \mathbf{c}_{j10} $, for $j=1, \ldots, 6$ from the assumed prior. Rather, we set the group allocations manually as in Figure~\ref{fig:sim_study:clust_assign}, in order to explore a wide spectrum of time-varying local clustering patterns. For example, for the age classes associated with the spline bases $3$ (top-right panel) and $4$ (bottom-left panel) the synthetic countries are grouped into stable clusters across time, albeit with differing co-clustering patterns. Conversely the remaining spline bases exhibit more complex dynamic clustering patterns (see, e.g. the spline basis $6$ where the synthetic countries often change cluster membership). 

Leveraging these simulated data, we perform Bayesian inference  under the model proposed in Section~\ref{section:model_formulation}, setting diffuse hyperparameters $a_\sigma = b_\sigma = a_\delta = b_\delta = a_\omega = b_\omega = 10^{-3}$, $a_M = 2 \cdot 10^{-3}$, $b_M = 10^{-3}$ and $a_{\alpha}=b_{\alpha}=1$. Consistent with standard practice in the Gaussian process literature \citep[][]{williams2006gaussian}, the entries of $\boldsymbol{\Sigma}$ in \eqref{eq:model_beta} are defined through a squared-exponential kernel having length scale $1.5$, i.e. $\boldsymbol{\Sigma}_{[t,t']}=\exp[-0.5(t-t')^2/(1.5)^2]$, for every $(t,t')$. Finally, to achieve an improved calibration of the proposed model, the mean vectors $\boldsymbol{\mu}_j$ in~\eqref{eq:model_beta} are defined in a data-driven manner. This is accomplished by first obtaining, separately for each $t=1, \ldots, 10$, an OLS estimate of the spline coefficients under model \eqref{eq_1} applied to data $\log m_{ixt}$, $i\,{=}\,1, \ldots, 5$, $\mathcal{X} =\{0, 1, \ldots, 100 \}$, with $f_{it}(x)$ as in \eqref{eq:f_it}. For each $j=1, \ldots, 6$, the resulting coefficients estimates  are subsequently smoothed via LOESS in order to obtain the desired data-driven specification for the entries $\mu_{j1}, \ldots, \mu_{j10}$ of $\boldsymbol{\mu}_j$. See Section~\ref{sec_42} for further details on hyperparameter settings and sensitivity of the results to these choices.

Under the above settings, posterior inference for the proposed model proceeds via Monte Carlo as outlined in Section~\ref{sec:estimation}, leveraging the samples produced by the Gibbs routine derived within Section~\ref{sec:gibbs}. Such a routine is run for $20{,}000$ iterations, discarding the first $10{,}000$ as a conservative burn-in. Traceplots and autocorrelation plots indicate satisfactory mixing of the chains. In fact, in this simulation study we observe  convergence much before the  burn-in employed. Nonetheless, we opted for a more conservative setting that can be considered as a default in general contexts, including in the motivating mortality data application within Section~\ref{sec_5}.

\begin{table}[t]
\centering
\caption{\footnotesize Simulation study. Posterior means of co-clustering accuracies for each combination $(j,t)$.}
\begin{tabular}{lccccccccccc}
 \toprule
 & $t=1$ & $t=2$ & $t=3$ & $t=4$ & $t=5$ & $t=6$ & $t=7$ & $t=8$ & $t=9$ & $t=10$ \\
\midrule
Spline 1 $(j=1)$ & 0.992 & 0.999 & 0.997 & 0.985 & 0.945 & 1.000 & 1.000 & 1.000 & 1.000 & 0.999 \\
Spline 2 $(j=2)$ & 1.000 & 1.000 & 1.000 & 1.000 & 1.000 & 0.992 & 1.000 & 1.000 & 1.000 & 0.999 \\
Spline 3 $(j=3)$ & 1.000 & 1.000 & 1.000 & 1.000 & 1.000 & 1.000 & 1.000 & 1.000 & 1.000 & 1.000 \\
Spline 4 $(j=4)$ & 1.000 & 1.000 & 1.000 & 1.000 & 1.000 & 1.000 & 0.999 & 1.000 & 0.998 & 1.000 \\
Spline 5 $(j=5)$ & 0.999 & 1.000 & 1.000 & 0.998 & 0.990 & 0.997 & 1.000 & 1.000 & 1.000 & 0.998 \\
Spline 6 $(j=6)$ & 0.949 & 0.937 & 0.987 & 0.985 & 0.994 & 0.994 & 0.991 & 0.988 & 0.997 & 0.998 \\
\bottomrule
\end{tabular}
\label{tab:sim_study:accuracy}
\end{table}

\begin{figure}[t]
    \centering
    \includegraphics[width =0.95 \linewidth]{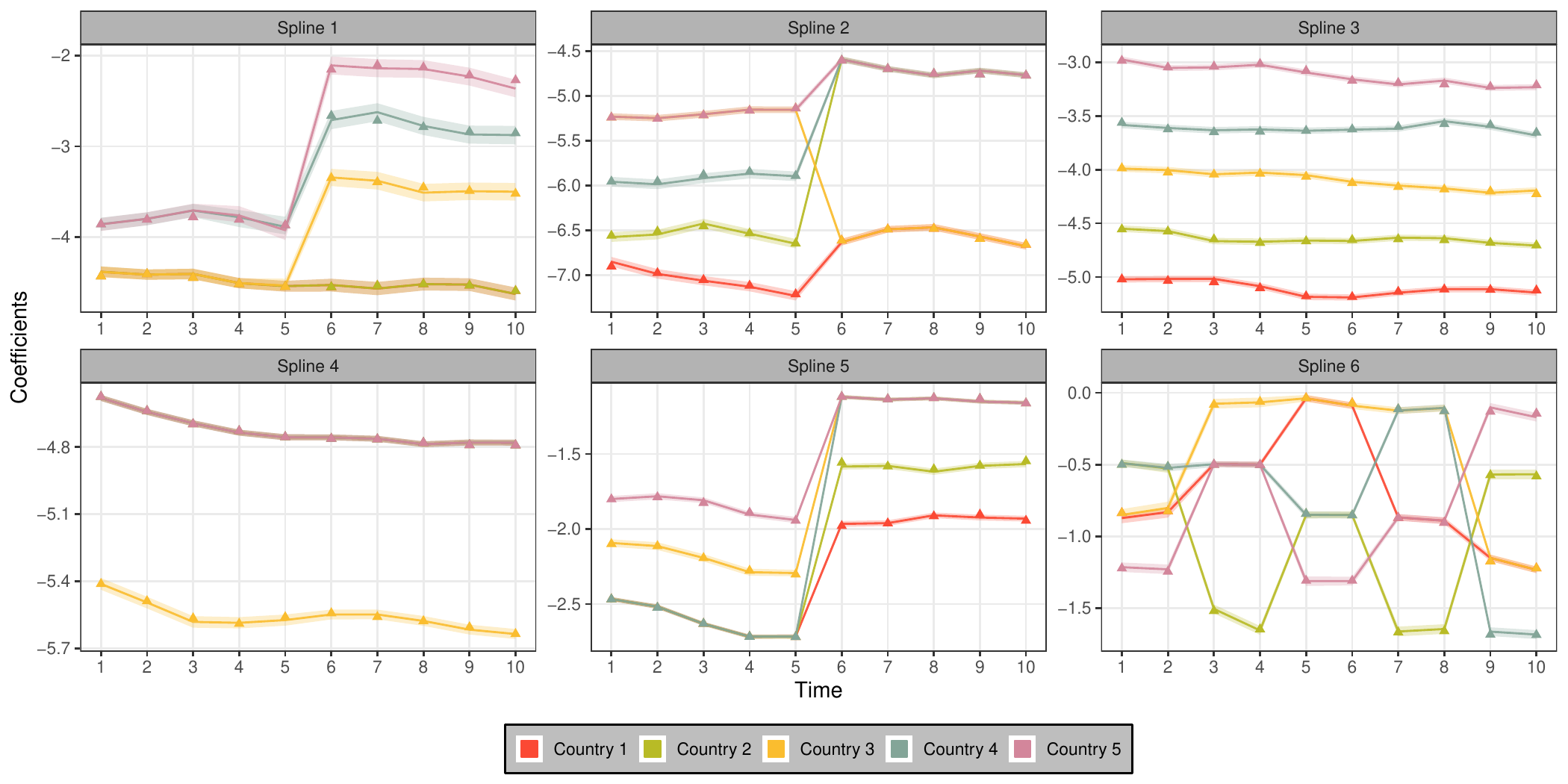}
    \caption{\footnotesize Simulation study. Posterior means of $\beta_{i j 1}, \ldots, \beta_{i j 10}$ (lines) and true values (points) for each country $i=1, \ldots, 5$ (colors) and basis function $g_j(x)$, $j=1, \ldots, 6$ (panels). Colored bands denote 95\% credible intervals.}
    \label{fig:sim_study:beta_units}
\end{figure}

Consistent with our overarching focus, we first assess in Table~\ref{tab:sim_study:accuracy} to what extent the proposed model is able to learn the {\em true} clustering structures displayed in Figure~\ref{fig:sim_study:clust_assign}  among the $5$ synthetic countries. To provide a comprehensive assessment for the clustering accuracy achieved by the posterior for $\mathbf{c}_{j1}, \ldots, \mathbf{c}_{j10} $,  $j=1, \ldots, 6$, beyond the one obtained under a single point estimate, we compute, for every Gibbs sample of $\mathbf{c}_{jt}$, $t=1, \ldots, 10$, $j=1, \ldots, 6$, the percentage of pairs of countries that are correctly co-clustered. The posterior means of these percentages over the $10{,}000$ retained samples are reported within  Table~\ref{tab:sim_study:accuracy}, and confirm the excellent performance of the proposed model in recovering the co-clustering patterns induced by the  {\em true}  group structures in Figure~\ref{fig:sim_study:clust_assign}. In particular, the accuracy measures are above $0.94$ for all the spline bases and time points, except for the $6$-th one where we observe a slight performance deterioration.  This result is expected since such a basis presents the most complex clustering patterns, with frequent changes of group memberships for all the countries. Nonetheless, even in this challenging regime, we still observe remarkable performance with all accuracy measures across time points above $0.93$.

As shown within Figure~\ref{fig:sim_study:beta_units}, the above remarkable performance in learning local clustering structures directly translates into highly accurate point estimates for the country-specific coefficient trajectories $\beta_{i j 1}=\beta^\star_{c_{ij1} j 1}, \ldots, \beta_{i j 10}=\beta^\star_{c_{ij10} j 10}$, $i=1, \ldots, 5$, $j=1, \ldots, 6$. These point estimates are obtained as detailed in Section~\ref{sec:estimation}, and compared with the associated {\em true} values within Figure~\ref{fig:sim_study:beta_units}. Results confirm the ability of the model to characterize the underlying trajectories of each spline coefficient accurately, even for complex underlying temporal dynamics. Under \eqref{eq_1}--\eqref{eq:f_clust_beta_p}, this implies effective learning of the data-generative mechanism for the synthetic log-mortality rates.

\subsection{\large 4.2 Second scenario}\label{sec_42}
In addition to the analysis presented in Section~\ref{sec_41}, the following section evaluates the robustness of the proposed model in a more realistic setting that emulates several structural features of mortality.  Consistent with this goal, we generate the log-mortality rates under a popular demographic model different from the one we designed, and assess the empirical performance of our approach within such a misspecified scenario. More specifically, raw data are downloaded from the Human Mortality Database \citep{hmd}, considering the mortality rates for $n=4$ countries (Italy, Sweden, the UK and USA) observed over $T = 10$ periods (from 2005 to 2014) and $5$-year age intervals.  For each country-period combination, separate Siler models \citep{siler1979} are fitted using the \texttt{MortalityLaws} package \citep{mortalitylaws-package} and used to construct mortality rates at ages $x \in \{0,\ldots,99\}$. In order to emphasize cross-country similarities, we enforce overlap among the mortality curves for manually selected combinations of countries, ages and periods. These overlaps are not induced by clustering among the coefficients of a spline-based decomposition of the mortality curves, and therefore it is of interest to assess whether the proposed model can still achieve accurate fit and obtain a sensible reconstruction of the underlying grouping structures.

Posterior inference in this misspecified scenario proceeds under the same default hyperparameters settings as those considered in Section~\ref{sec_41}, except for the choice of the B-spline bases that are defined as in Figure~\ref{fig:splines} to assess in practice the basis specification considered in the application in Section~\ref{sec_5}. We also follow the same data-driven strategy outlined in Section~\ref{sec_41} to define the elements of $\boldsymbol{\mu}_j$ for each basis $j=1,\ldots,20$.

As a first performance assessment  let us focus on the quality of the clustering structures inferred under the proposed model. Since in this misspecified setting there is no ``true'' partition structure for the spline coefficients, we evaluate the coherence between the co-clustering probabilities resulting from our model and the overlaps in the observed mortality rates. Specifically, recalling the discussion within Section~\ref{subsection:model_formulation}, it is sensible to expect that, even in misspecified settings, large co-clustering probabilities for a generic pair of countries in the coefficient of a given spline basis would correspond to similar mortality rates for that pair over the age interval in which the basis is {\em leading}. Empirical evidence in favor of such an intuition would not only clarify the robustness of the proposed model across several data-generating mechanisms, but also reassure on the fact that the co-clustering at the level of the spline coefficients translates into local similarities in the mortality surfaces over the age range where the spline is  {\em leading}.  To quantify this behavior, for each pair of countries $i$ and $i^\prime$, spline basis $j=1, \ldots, p$ and period $t=1, \ldots, T$, we compare the posterior co-clustering probability $\hat{\bf P}_{jt[i, i^\prime]}$ with the weighted average of the absolute differences among log-mortality rates $\tau_{jt[i,i']} = \sum_{x \in \mathcal{I}_j} ([g_j(x)/\sum_{z \in \mathcal{I}_j} g_j(z)] {\cdot} |\log m_{ixt} - \log m_{i^\prime xt}|)$, over the age interval $\mathcal{I}_j$ where the $j$-th spline basis is {\em leading}.  Table~\ref{tab:median_variance} summarizes the results of this comparison by displaying the median of the distances $\tau_{jt[i,i']}$ for the pairs of countries with low ($\leq 0.5$) or high ($>0.5$) posterior co-clustering probabilities. The results, aggregated across periods and spline bases (stratified within three macro age groups),  indicate that the proposed model identifies meaningful grouping structures across all macro age groups, with high co-clustering probabilities achieved for pairs of countries that display highly similar mortality rates. Notice that the median and the standard deviation of all the $\tau_{jt[i,i']}$ within the three macro age classes correspond to $0.346\,(0.29)$, $0.406\,(0.27)$ and $0.122\,(0.15)$, respectively. Hence, pairwise distances in the order of $10^{-2}$ imply a close similarity among the observed rates of co-clustering countries, thus supporting the practical meaningfulness  of the proposed model and clustering procedures.
 
The above results proved also robust to moderate changes in the  hyperparameter values. More specifically, varying  $a_\sigma, b_\sigma, a_\delta, b_\delta, a_\omega, b_\omega$ in the range $\{10^{-2}, 10^{-1}, 1\}$ does not substantially affect the results obtained in  Table~\ref{tab:median_variance} under the non-informative default setting $10^{-3}$, which allows to weight more in the posterior the information provided by the data. Under these alternative specifications, all variations  in  Table~\ref{tab:median_variance} are at most by $0.05$, thereby mitigating sensitivity concerns with respect to the choice of $a_\sigma, b_\sigma, a_\delta, b_\delta, a_\omega,$ and  $b_\omega$. To provide a comprehensive assessment, we also evaluated a non-informative specification for $\boldsymbol{\mu}_j$ that fixes its elements to zero, rather than leveraging the proposed data-driven specification. In this case, the results in  Table~\ref{tab:median_variance} display still  limited, although more evident, variations of $\approx$ 0.1--0.2 that point towards a slightly deteriorated clustering performance under model misspecification. This finding indicates that a data-driven specification for $\boldsymbol{\mu}_j$ is expected to provide improved  fit, and hence, should be preferred in applications. In fact, centering $\boldsymbol{\mu}_j$ around zero induces  a prior on the log-mortality rates that is also centered at zero, and hence, unrealistic for mortality applications.
 
To complete the analysis, we further assess the ability of the proposed model to characterize the underlying data-generative mechanism by evaluating its performance in out-of-sample prediction.  To this end, we consider four settings in which out-of-sample observations are selected uniformly across countries, ages, and periods under growing proportions  $20\%$, $30\%$, $40\%$ and $50\%$, respectively. In addition to the correctly-specified  \citet{siler1979} model, we consider as competitors for this assessment also other popular parametric models from the demographic literature, namely,  \citet{carriere1992}, \citet{heligman1980},  and \citet{martinelle1987}, as implemented in \citet{mortalitylaws-package}. Since these models are designed for mortality rates observed in a given period for a single population, each specification is fitted separately for every combination of country and period. The comparison among models is  in terms of root mean squared error (RMSE) between observed and predicted log-mortality rates. The parametric models provide predictions for such rates at all ages, allowing a direct evaluation of model performance on out-of-sample data. Conversely, under our specification, out-of-sample data are treated as missing values and sampled at each iteration of the Gibbs sampler from their corresponding posterior predictive distribution. At convergence, point estimates for such observations are obtained via posterior means.

\begin{table}[t]
\centering
\caption{\footnotesize Simulation study. Median of the distances $\tau_{jt[i,i']}$ between log-mortality rates for pairs of countries  with estimated co-clustering probabilities below or above $0.5$. Results are aggregated across periods and spline bases (stratified in three macro age groups).}
\begin{tabular}{lcccccc}
\toprule
 & \multicolumn{2}{c}{Early mortality}  
 & \multicolumn{2}{c}{Adult mortality} 
 & \multicolumn{2}{c}{Senescent mortality} \\
\cmidrule(lr){2-3} \cmidrule(lr){4-5} \cmidrule(lr){6-7}
 & $\leq 0.5$ & $> 0.5$ 
 & $\leq 0.5$ & $> 0.5$ 
 & $\leq 0.5$ & $> 0.5$ \\
\midrule
 & 0.3920  & 0.0551 
 & 0.4590 & 0.0313 
 & 0.1620 & 0.0399 \\
\bottomrule
\end{tabular}
\label{tab:median_variance}
\end{table}

\begin{table}[t]
\centering
\caption{\footnotesize Simulation study. Predictive performance  of the different models under analysis for varying out-of-sample proportions. Average RMSE (and standard deviation) across 50 replications.}
\begin{tabular}{c|c|cccc}
\toprule
Out-of-sample \% & \textsc{Proposed model} & \textsc{Siler} & \textsc{Heligman-Pollard} & \textsc{Carriere} & \textsc{Martinelle}  \\
\midrule
20\% & 0.10 (0.03) & 0.12 (0.04) & 0.16 (0.06) & 0.22 (0.06) & 0.28 (0.02) \\
30\% & 0.11 (0.03) & 0.13 (0.03) & 0.18 (0.04) & 0.22 (0.04) & 0.29 (0.02) \\
40\% & 0.13 (0.04) & 0.16 (0.03) & 0.20 (0.04) & 0.22 (0.04) & 0.30 (0.01) \\
50\% & 0.13 (0.03) & 0.16 (0.03) & 0.24 (0.07) & 0.22 (0.03) & 0.31 (0.01) \\ 
\bottomrule
\end{tabular}
\label{tab:oos_metrics}
\end{table}

Table~\ref{tab:oos_metrics} reports results averaged across $50$ replicated experiments. Although misspecified, the proposed formulation displays a predictive performance competitive with the correctly-specified   \citet{siler1979} model, and superior to those of other popular competitors routinely employed in the demographic literature \citep[i.e.][]{carriere1992,heligman1980,martinelle1987}. This finding clarifies that the B-spline bases in Figure~\ref{fig:splines}, combined with the structured cluster-oriented prior on the associated coefficients, achieve an effective balance between flexibility and parsimony/interpretability in modeling the complex data-generative mechanisms underlying mortality data observed across diverse countries. These benefits, along with the   richer  inferential output offered by the proposed model, are further showcased in the real-data application in Section  \ref{sec_5}.

\section{\large 5. Local clustering of age-period mortality surfaces for 14 countries} \label{sec_5}
We conclude by illustrating the performance of the proposed Bayesian model in learning the local clustering structures across ages and periods induced by the log-mortality rates of $14$ countries (Australia, Belgium, Canada, Switzerland, Denmark, Spain, Finland, France, Italy, the Netherlands, Norway, Sweden, the United Kingdom and the United States). The original data $\log m_{ixt}$ are retrieved from the Human Mortality Database \citep{hmd} for ages from $0$ until $98$ years-old, over a time horizon of $88$ calendar years (1933--2020).

In applying the proposed model to the above data we follow standard practice in multi-country studies \citep[e.g.][]{Aliverti2022,Li&Lee2005} by considering separate analyses for male and female sub-populations.  Posterior inference proceeds under the same hyperparameters and Gibbs settings as those considered for the simulation studies in Section~\ref{sec_42}, leveraging the B-spline bases displayed in Figure~\ref{fig:splines}. As in the simulations, the traceplots for the quantities of interest and the convergence diagnostics did not provide evidence against convergence, while sensitivity analyses, analogous to those presented in Section~\ref{sec_42},  confirm the robustness of the model with respect to moderate variations in the hyperparameter values also in the applications. For instance, for the male population, the median distances between the log-mortality rates for  pairs of countries with low  co-clustering probability range over $[0.325, 0.328]$, $[0.336, 0.339]$, and $[0.170, 0.180]$ for early, adult, and senescent mortality, respectively. Correspondingly, the median distances for pairs with a high probability of co-clustering range over $[0.140, 0.143]$, $[0.142, 0.143]$, and $[0.071, 0.072]$. When compared to the median and the standard deviation of all distances $\tau_{jt[i,i']}$ across the same three age  groups --- equal to $0.290$ ($0.25$), $0.286$ ($0.29$), and $0.137$ ($0.15$) --- these results confirm the reliability already demonstrated in the simulation study. As a further evidence of the quality and robustness of the proposed model, we assess its fit across the different hyperparameter settings via the relative median absolute error, defined as the absolute difference among the estimated and observed log-mortality rates divided by the observed absolute rates. For the male population, the median of this metric across all combinations of countries, ages, and periods is $0.0090$, $0.0090$, $0.0091$, and $0.0092$ for increasing values of the hyperparameters   $a_\sigma$, $b_\sigma$, $a_\delta$, $b_\delta$, $a_\omega$, $b_\omega$  from $10^{-3}$ to $1$. Similar robustness and accurate fit is observed also for the female population.

Section~\ref{sec_51} summarizes the results of posterior inference under the proposed model, whereas Section~\ref{sec_52} highlights previously unexplored local clustering structures with a specific focus on the United States. Finally, Section~\ref{sec_53} explores possible associations between the novel co-clustering patterns inferred by the proposed  model and relevant socio-economic variables, including gross domestic product (GDP) and health expenses.

\subsection{\large 5.1 Results}
\label{sec_51}
\begin{figure}[t]
    \centering
    \includegraphics[width =0.95 \linewidth]{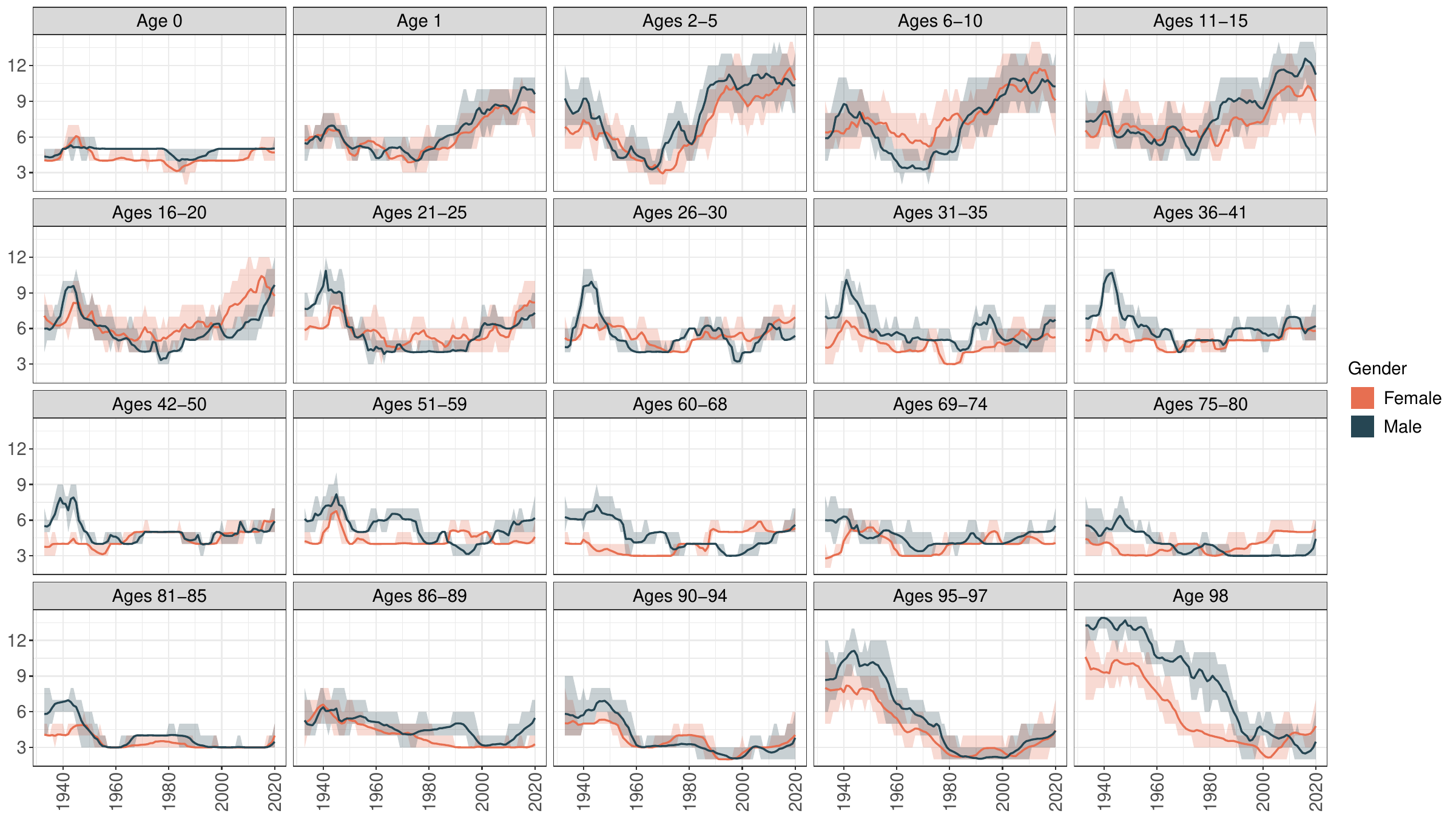}
    \caption{\footnotesize For each spline $j=1, \ldots, 20$ (denoted here through the age range where such a spline is {\em leading}) time trajectories for the posterior mean of the number of clusters induced at the coefficient level. Colored bands correspond to period-specific 95\% credible intervals.}
    \label{fig:real_data:comparison:avg_num_clus}
\end{figure}

Figure~\ref{fig:real_data:comparison:avg_num_clus} displays the evolution of the posterior means for the number of clusters inferred by the proposed model across the age intervals associated with the $20$ spline bases within Figure~\ref{fig:splines}, over the temporal window analyzed. Such a quantity is generally stable for both infant mortality (age $0$) and adult/elder mortality (from age $51$ until $89$). Instead, an increased variability in the number of clusters is observed for children and adolescents (from age $1$ to $25$) and late mortality (from age $90$), with different patterns. Besides supporting the need for allowing the group structures among countries to vary locally with ages and periods, these trends can be interpreted as a measure of variability in mortality rates across countries, with larger number of groups corresponding to age intervals where the different countries display higher temporal dissimilarities. Consistent with this interpretation,  Figure~\ref{fig:real_data:comparison:avg_num_clus} suggests that senescent mortality is characterized by an increasing level of similarity among countries over the recent years, whereas child mortality exhibits an opposite trend after the early 1970s, indicating a divergence among countries that might have been driven by social phenomena such as the baby boom. Overall, males and females are characterized by similar patterns, with few notable exceptions. The number of clusters for eldest ages converges to few groups for both sub-populations, although the males are generally characterized by higher dissimilarities among countries since the first period under investigation. It is also interesting to notice how the mortality of young and adult males is affected by World War II in the $1940$s, and reaches a plateau only after the peak associated with the military services.

\begin{figure}[t]
    \centering
    \includegraphics[width = 0.8\linewidth]{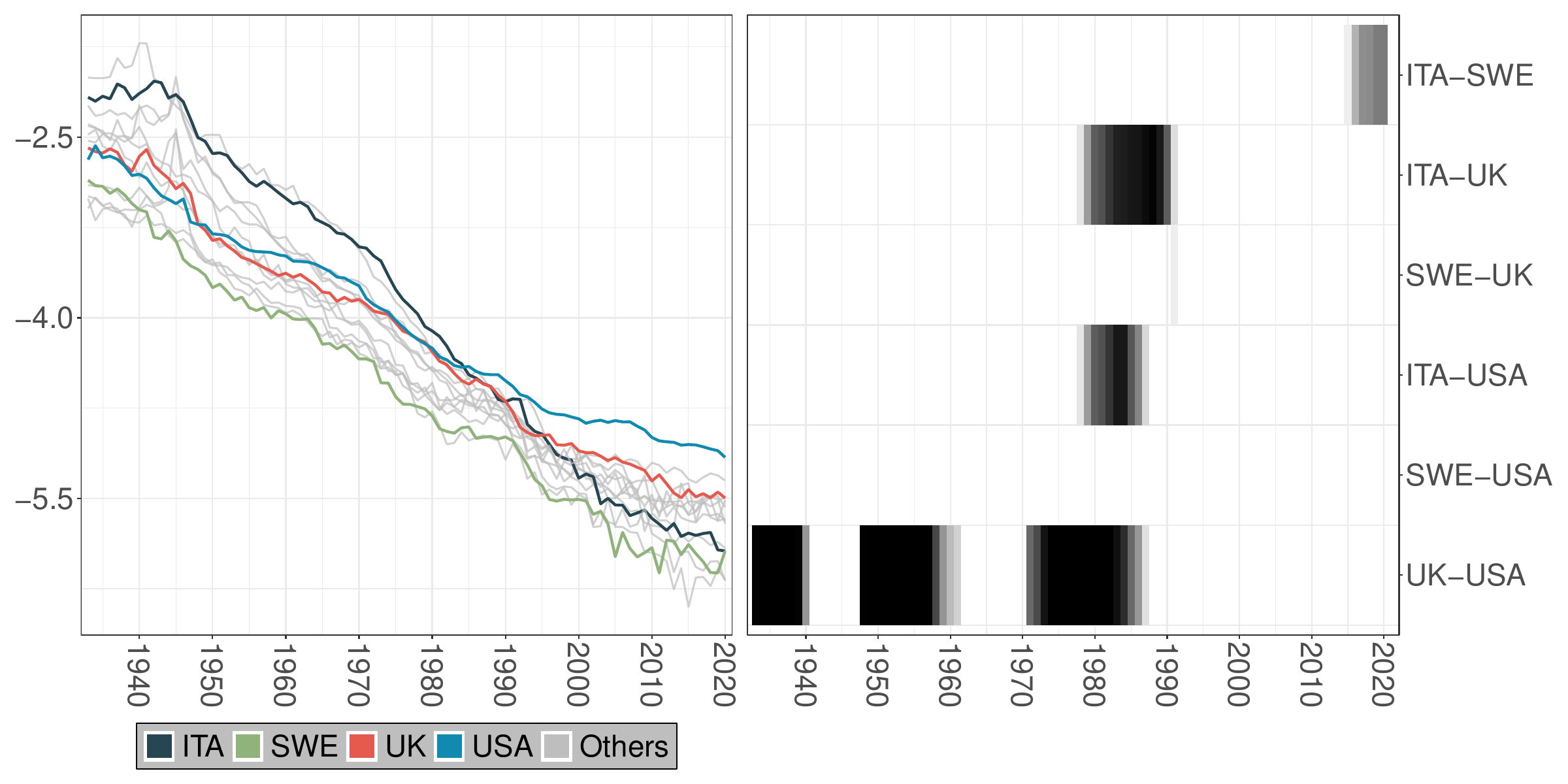}
    \caption{\footnotesize Observed male infant log-mortality rates for Italy (ITA), Sweden (SWE), the United Kingdom (UK) and the United States (USA) (left), and estimated dynamic co-clustering probabilities for the corresponding pairs with respect to the coefficient of the {\em leading} spline at age $0$ (right). Black denotes high co-clustering probability.}
    \label{fig:real_data:men:prob_and_data}
\end{figure}

The ability of the proposed model to effectively capture local clustering patterns is illustrated in Figure~\ref{fig:real_data:men:prob_and_data}, where the observed infant log-mortality rates in Italy, Sweden, the UK and the USA are compared with the estimated probabilities of co-clustering with respect to the coefficients of the {\em leading} spline at age $0$. This specialized analysis is useful to further illustrate the main advantages of the proposed approach and its implications for the global characterization of mortality rates. In particular, the right panel of Figure~\ref{fig:real_data:men:prob_and_data} indicates a fluctuating probability of co-clustering between the UK and the USA (bottom row), with larger values before $1940$, in the early $1950$s, and more recently from $1970$ to $1985$, in agreement with the observed trends of mortality rates reported in the left panel.  Evidence of co-clustering is observed also for Italy with both the UK and the USA around the  $1980$s.  In contrast, Sweden is characterized by a peculiar and separate trajectory, that  progressively overlaps with Italy  only in recent years (top row).

\begin{figure}[t]
    \centering
    \includegraphics[width = 0.87\textwidth]{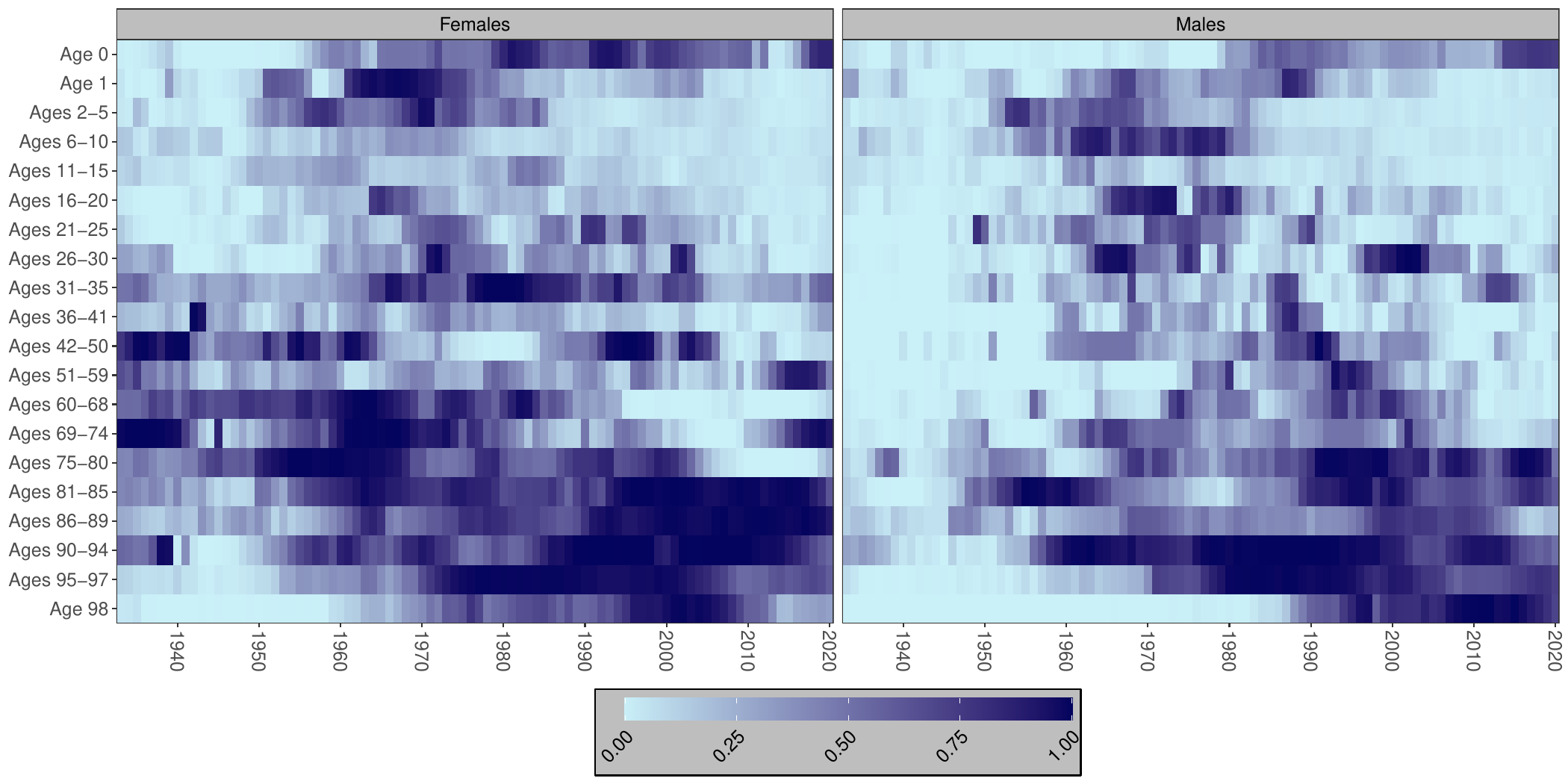}
    \caption{\footnotesize Third quartile of the estimated co-clustering probabilities among pairs of countries, across all combinations of periods (columns) and spline bases  (denoted in the rows through the age range where each spline is {\em leading}).}
    \label{fig:real_data:heatmap}
\end{figure}

To fully explore the evolution of local co-clustering patterns, beyond the above specialized analysis, Figure~\ref{fig:real_data:heatmap} summarizes the estimated co-clustering probabilities across all pairs of countries for every combination of age interval where each spline is {\em leading} (rows) and periods (columns). Specifically, the co-clustering probabilities are summarized via the  third quartiles across all pairs of countries.  Thus, large values for this quartile imply that at least $25\%$ of country pairs present a high  co-clustering probability, while low values indicate higher heterogeneity in country-specific patterns at given combinations of ages and periods.  Consistent with Figure~\ref{fig:real_data:comparison:avg_num_clus}, the results in Figure~\ref{fig:real_data:heatmap} show that evidence of increasing co-clustering is progressively more present both as the time advances and as  population ages, demonstrating a dual-directional reinforcement of these similarities.  For instance, in the splines associated with the oldest age groups of the male sub-population (right panel of Figure~\ref{fig:real_data:heatmap}), the third quartile of the co-clustering probabilities has increased steadily in recent years. This compression effect is more pronounced at advanced ages, whereas younger age classes display systematically lower values for the  third quartile than older ones. These findings are consistent with previous evidence on demographic convergence \citep[e.g.][]{vaupel2011,wilson2011}. A similar pattern is observed in the female sub-population, although the upward trend in  probabilities of co-clustering appears to have started earlier, suggesting an anticipated convergence process among females.

\begin{figure}[t]
    \centering
    \includegraphics[width=0.9\textwidth]{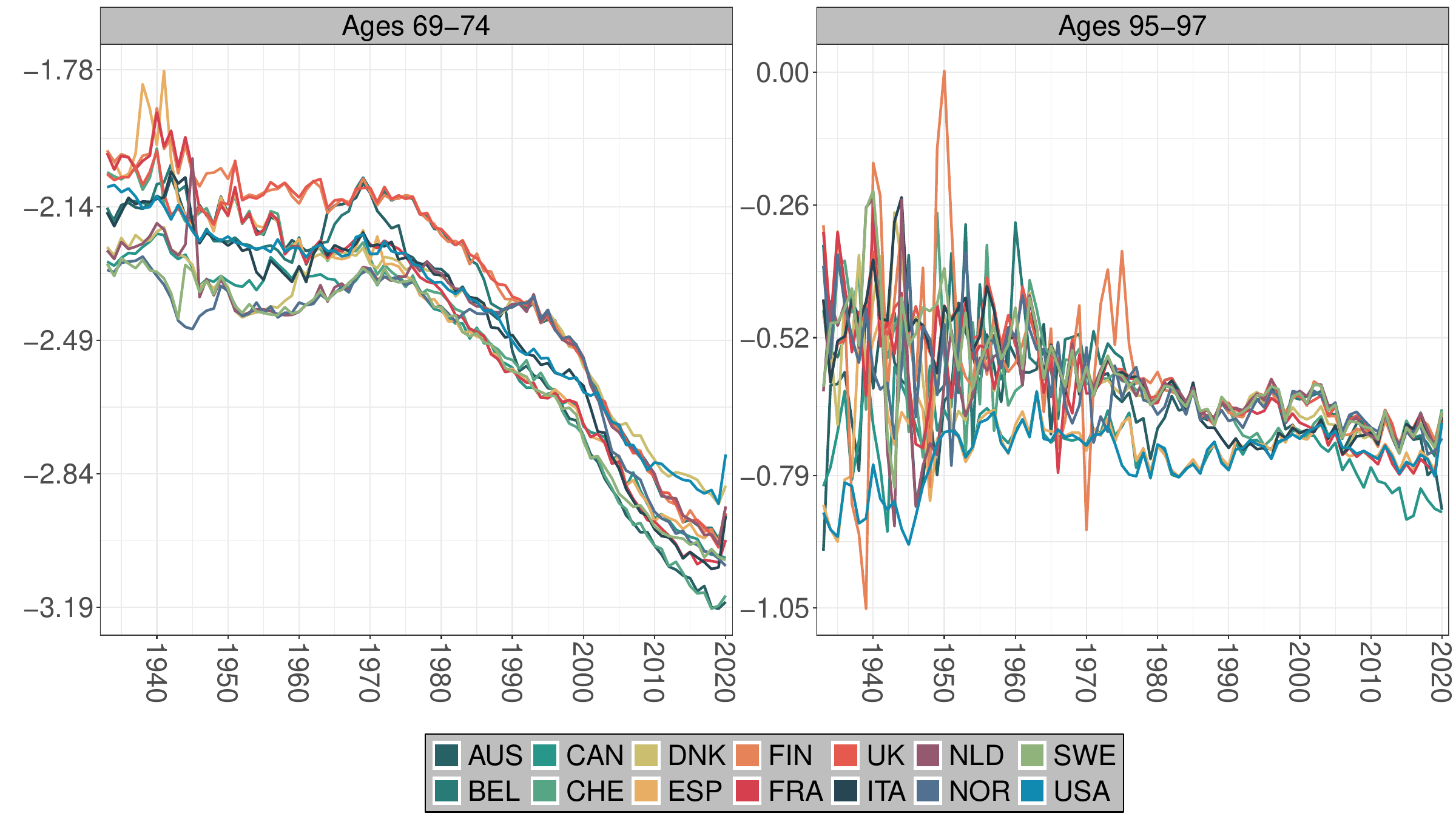}
    \caption{\footnotesize Estimated trajectories of the coefficients $\beta_{i,14,t}$ and $\beta_{i,19,t}$ for the different countries in the male population. The two coefficients correspond to the spline bases {\em leading} in the age intervals 69--74 and 95--97, respectively.}
    \label{fig:real_data:men:beta_units_spline_14_and_19}
\end{figure}

Focusing on child mortality, the younger age classes (from age $1$ to $20$) show a neat temporal pattern, with most of the co-clustering concentrated between the late 1950s and the early 1980s.  This behavior is coherent with the evolution of the number of clusters in Figure~\ref{fig:real_data:comparison:avg_num_clus},  and suggests that the countries under investigation have been characterized by different levels of child mortality until World War II, followed by a rapid improvement of mortality rates in the early $1960$s that created few common clusters with similar mortality patterns.  After the $1980$s, the co-clustering structures become more irregular and countries deviate to more individual trends. This finding is worth future investigations. The co-clustering patterns among countries for adult mortality are reported across the middle rows of Figure~\ref{fig:real_data:heatmap}. These show, in general, a common diagonal structure compatible with cohort effects, combined with a more marked variability across consecutive periods in the third quartile of the co-clustering probabilities. This latter phenomenon can be more clearly appreciated, for the male sub-population, in the left panel of Figure~\ref{fig:real_data:men:beta_units_spline_14_and_19}, that reports the estimated trajectories for the country-specific coefficients $\beta_{i,14,t}$ of the spline basis which is {\em leading} in the age interval 69--74. Such  trajectories show how the Netherlands and Denmark are characterized by lower mortality rates than several countries until $1960$, while Belgium, Finland and the UK experienced some of the highest rates during that period.  Later on, mortality  generally drops for all countries, while the Netherlands and Denmark experience a relative increment until $1970$ that makes them co-cluster with higher mortality countries, after which they jointly follow a common decreasing trend. Structural changes in cross-country relations characterize this period.  The potential impact of such changes is visible within the right panel of Figure~\ref{fig:real_data:men:beta_units_spline_14_and_19}, which displays the evolution of the country-specific coefficients $\beta_{i,19,t}$ associated with the spline basis that is {\em leading} at ages 95--97. These trajectories  illustrate how after the $1970$s mortality levels of elder ages tend to collapse into fewer clusters with stable trends for several years.

To further study the similarity between the group structures among countries inferred for the male and female log-mortality rates, we also computed the normalized variation of information  (NVI) \citep[][]{Wade&Ghahramani2018} between the allocation vectors estimated for these two sub-populations. Overall, the similarity among male and female partitions is more evident for the two youngest age classes ($0$ and $1$), with NVI values below 0.44 and 0.58, respectively, over the entire time window. For example, female infant mortality in 2020 is divided into four groups. The first comprises the  USA, the second Canada, the third Central European countries (Belgium, Switzerland, Denmark, France, the Netherlands and the UK) and Australia, and the fourth encodes both Mediterranean Europe (Italy, Spain) and Nordic countries (Finland, Norway, Sweden). The corresponding partition for the male sub-population consists, instead, of five groups; the first three are the same as those of the female sub-population. The last two are obtained by splitting the fourth cluster of female population into two groups, containing Italy, Spain, Norway and Finland, Sweden, respectively. More remarkable differences between male and female sub-populations are observed among middle-aged individuals, with NVI values above 0.5 after the 1960s. Interestingly, for the elderly population, a trend of increasing similarity started in the late 1980s, mirroring the patterns of  young ages. These trends reflect the progressive convergence of mortality  across countries \citep{vaupel2011} for both males and females.

\subsection{\large 5.2 A specialized focus on the United States}
\label{sec_52}
We devote here special attention to the analysis of the co-clustering probabilities between the USA and the other countries, estimated under the proposed model. This specialized focus is motivated by the fact that the USA has exhibited peculiar patterns in the recent years, with rising mortality rates among certain demographic groups and a decline of life expectancy \citep[e.g.][]{bergeron2020,case2021deaths,glei2022us}. Such a phenomenon has been extensively studied in the literature, with a general consensus attributing these unexpected mortality increments to factors such as persistent disparities in healthcare access, increasing suicide rates and, more recently, to the opioid epidemic \citep[see, e.g.][]{woolf2019us}.  To complement and extend these findings, it is therefore particularly interesting to study how the USA co-clusters with the other countries under investigation in terms of mortality rates.

\begin{figure}[t]
    \centering
\begin{subfigure}{\textwidth}
    \centering
    \includegraphics[width = 0.95\textwidth]{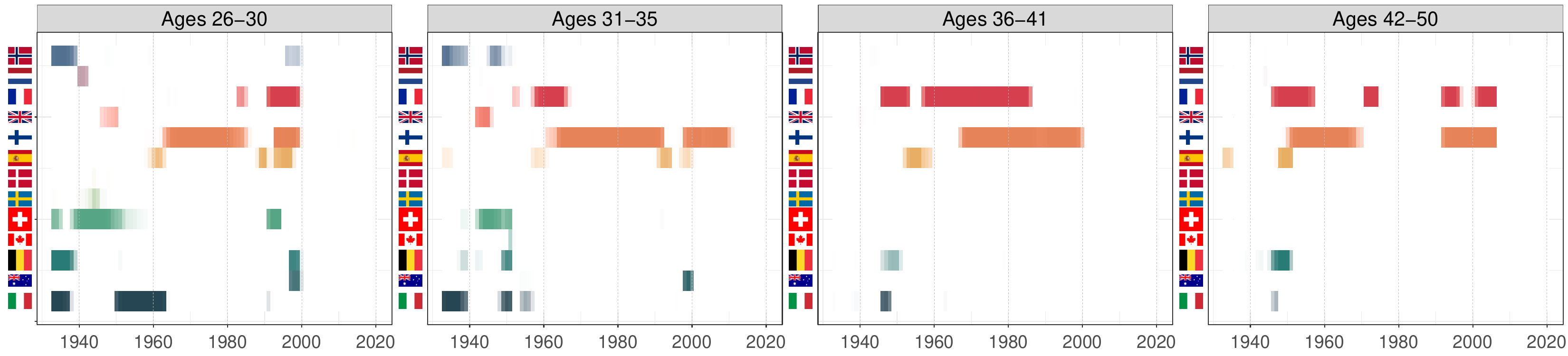}
    \caption{\footnotesize Male population}
    \label{fig:real_data:men:eta2:withUSA}
\end{subfigure}
\begin{subfigure}{\textwidth}
    \centering
    \includegraphics[width = 0.95\textwidth]{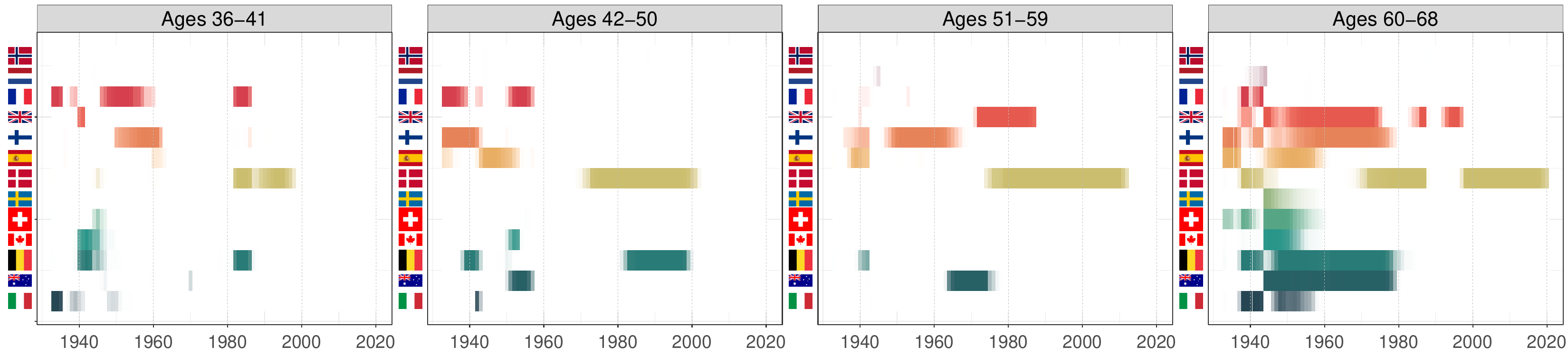}
    \caption{\footnotesize Female population}
    \label{fig:real_data:women:eta2:withUSA}
\end{subfigure}
    \caption{\footnotesize Dynamic  co-clustering probabilities between the USA and the other countries, for selected spline bases that are {\em leading} at age intervals of interest. Colors go from light to dark as the probability varies from low to high.}
\end{figure}

Consistent with the above goal, Figure~\ref{fig:real_data:men:eta2:withUSA} displays the dynamic probabilities of co-clustering among the USA and other countries for selected spline bases in the male sub-population, estimated under the proposed model. The results  in Figure~\ref{fig:real_data:men:eta2:withUSA}  point toward a persistent co-clustering among the USA and Finland. Such a finding indicates strong similarities in terms of premature male mortality and can be attributed to suicide incidence and cardiovascular diseases.  Indeed, although Nordic countries provide universal and publicly funded welfare systems for all citizens, significant disparities still persist across the socio-economic spectrum due to the so-called ``Nordic Paradox'' \citep[see, e.g.][]{hojstrup2023,mackenbach2017}. As a result, Figure~\ref{fig:real_data:men:eta2:withUSA} suggests that the ``Death of Despair'' phenomenon in the USA \citep[][]{case2021deaths,glei2022us} might have interesting similarities with the ``Nordic Paradox''.

Focusing on the female sub-population, Figure~\ref{fig:real_data:women:eta2:withUSA} provides evidence of a relatively-persistent co-clustering among the  USA and Denmark females at adult ages, varying across calendar years and with age intervals. Such a peculiar result  should be further investigated, since the interwar generation of Danish females represents a relevant demographic group that comprises cohorts who experienced a peculiar stagnation in life expectancy, resulting in a divergent trajectory compared to other Nordic countries \citep{lindahl2016pnas}. Therefore, recent improvements in life expectancy can be linked to a cohort effect associated with this group that is visible also from the co-clustering patterns in Figure~\ref{fig:real_data:women:eta2:withUSA}.  Interestingly, Belgian females join these clusters in specific age groups and time periods; a similar trend can be observed for  UK at age 51--59. Such patterns might be related to a higher proportion of diseases of the circulatory system for Belgian females \citep{bergeron2020} and to general declines in life expectancy  in the UK \citep{ho2018}.

\subsection{\large 5.3 Associations among local clustering structures and socio-economic indicators}
\label{sec_53}
The results in Sections~\ref{sec_51}--\ref{sec_52} showcase relevant co-clustering structures among countries. These structures display convergence phenomena over periods for specific age classes \citep[e.g.][]{oeppen2002}, along with evidence of growing disparities for other ages. Available studies suggest that such patterns are  driven by socio-economic factors such as quality of healthcare, education and life standards \citep[e.g.][]{EU_Commission2013,marmot2005,vallin2004}. 

To provide empirical support to the above findings, we conclude our analysis by assessing to what extent the grouping structures inferred by the proposed model are associated with relevant socio-economic indicators, covering, in particular, the per-capita gross domestic product (GDP) in USD, unemployment rate, health expenditure (as the percentage of a country's GDP), nutrition and child vaccination rates (against diphtheria, tetanus, pertussis); refer to  \citet{OECD_data_warehouse} for more details. Data on these indicators are available, for all countries, across varying time periods. The earliest data for the GDP date back to 1970, whereas child vaccination rates and health spending are available starting from $1990$. Finally, data on nutrition quality and unemployment rate start from $2010$. Following standard practice, the association between these socio-economic indicators and the clustering structures among countries inferred by the proposed model is evaluated using the $\eta^2 \in [0,1]$ coefficient, which quantifies the variability of each indicator between the inferred clusters with respect to the total variability of such an indicator. Values of the $\eta^2$ close to $1$ imply that the group structures among countries learned on the basis of similarities in the associated mortality rates are also explicative of the variability for the indicator under analysis, thereby suggesting possible associations. Notice that this measure does not inform on the direction of such an association, nor on possible causal interpretations. Nonetheless,  it offers a sensible perspective for prioritizing specific indicators within explanatory studies on the socio-economic determinants of multi-country mortality patterns.

\begin{figure}[t]
    \centering    
    \includegraphics[width = 0.92\textwidth]{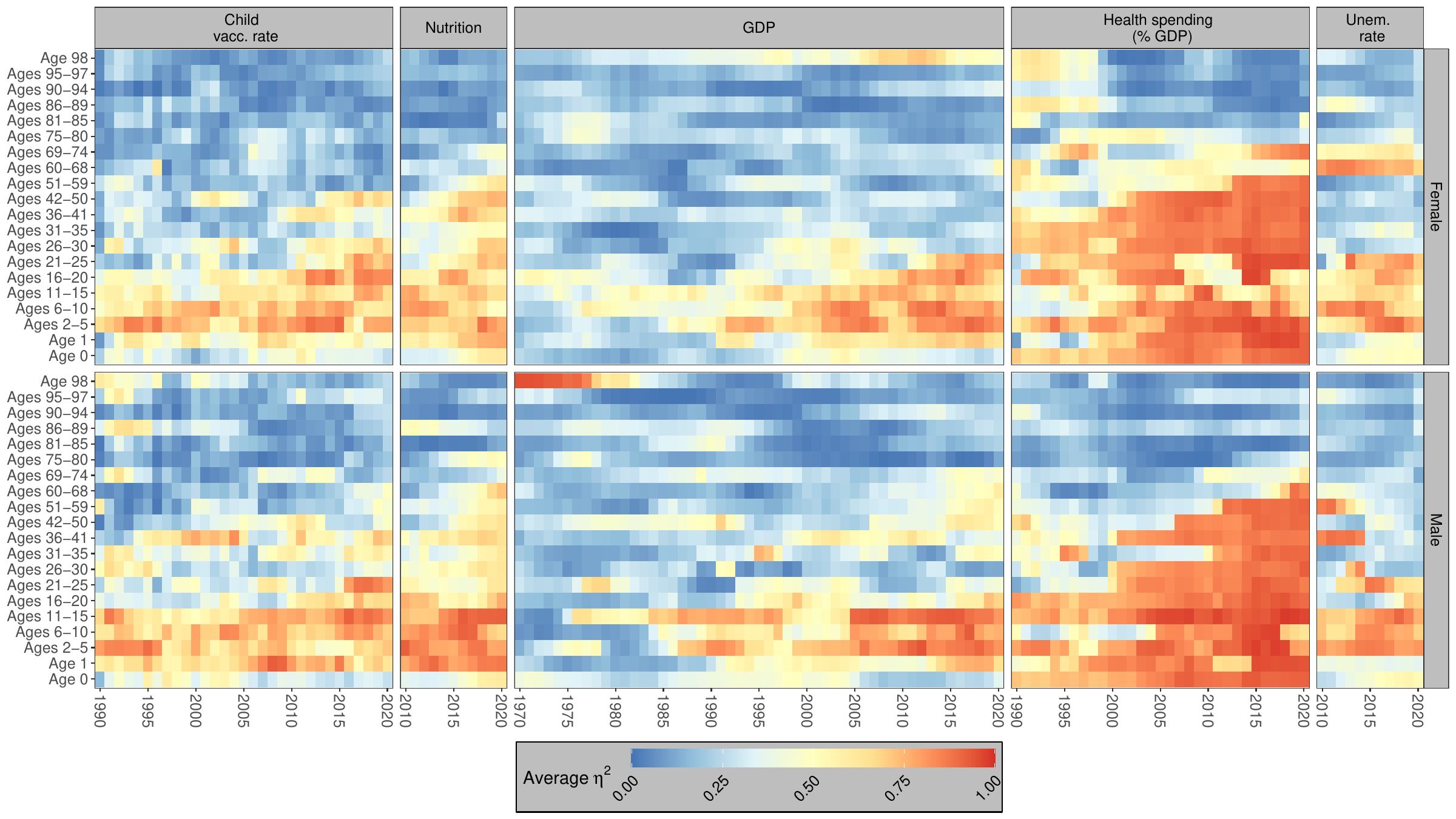}
    \caption{\footnotesize Posterior mean of $\eta^2$ coefficients between selected socio-economic indicators and the inferred  mortality-based clusterings in the female and male sub-populations. Results are shown for all combinations of periods (columns) and spline bases (denoted in the rows through the age range where each spline is {\em leading}).    }
    \label{fig:real_data:men:eta2:tiles}
\end{figure}

Figure~\ref{fig:real_data:men:eta2:tiles} illustrates, for both the female and male sub-population, the posterior means of the $\eta^2$ coefficients in matrix form. The rows correspond to the age intervals $j=1,\ldots, p$ associated with the different spline bases, columns to periods $t=1, \ldots, T$, and the panels to socio-economic indicators $q=1, \ldots, Q$. In order to obtain these posterior means we first compute the $\eta^2$ between each posterior sample of $\mathbf{c}_{jt}= [c_{1jt}, \dots, c_{njt}]$ and the vector ${\bf w}_{qt}=[w_{q1t}, \ldots, w_{qnt}]$ whose generic element $w_{qit}$ denotes the value of the socio-economic indicator $q$, for country $i$ in period $t$. This produces posterior samples of  the $\eta^2$ for every spline basis $j=1, \ldots, p$, period $t=1, \ldots, T$ and socio-economic indicator $q=1, \ldots, Q$. Averaging over these samples yields the posterior means in the matrices in Figure~\ref{fig:real_data:men:eta2:tiles}, which highlight a general concordance in the $\eta^2$ coefficients for the female and male sub-populations with non-trivial associations between the inferred clusters and the socio-economic indicators analyzed, particularly at young ages, and with different patterns across periods. For example, nutrition, GDP, health spending and the unemployment rate appear to be associated with mortality clusters through a lower-diagonal cohort structure spanning a large spectrum of ages, from childhood until young-adults and adults. This suggests that past incentives have had a lasting impact on specific cohorts, ultimately improving the lifespan of those who benefited from these initiatives. Not surprisingly, child vaccination rates and nutrition are generally less associated with clusters  at age $0$. In fact,  such interventions cannot prevent neonatal and infant mortality, which are caused by specific factors \citep[such as, birth defects, sudden infant death syndrome or accidents; see][]{eberstein1990,macdorman2013}. Instead, such indicators show a stronger association with the inferred clusters from age 1 until late adolescence, consistent with the protective effects of vaccination and proper nutrition in such ages.

\section{\large 6. Conclusions and future research directions}\label{sec_6}
Available statistical models for mortality data are either designed for analyzing single countries in isolation or for inferring global group structures among multiple countries with respect to the entire age-period mortality surface. This perspective prevents one from unveiling more nuanced similarity patterns that are observed in practice over specific combinations of ages and periods, thus limiting the possibility to learn and quantify relevant demographic phenomena localized at specific age classes and periods. 

In this article, we overcome these limitations through a multi-country model that characterizes the age pattern of mortality via a B-spline expansion, and incorporates both temporal and age-specific clustering structures by allowing the B-spline coefficients to change in time via separate temporal random partition priors with basis-specific Gaussian processes regulating the temporal evolution of such coefficients. The resulting Bayesian formulation is amenable to tractable posterior inference via a Gibbs sampling algorithm that facilitates  interpretable reconstruction of group structures among countries, varying locally with both ages and periods. The advantages of the newly-proposed model are illustrated in simulation studies and in an application to mortality data from 14 OECD countries, where our formulation unveils fundamental local clustering structures, including unexplored ones that motivate future research in the area. For example, the in-depth analysis of the USA clustering patterns  in Section~\ref{sec_52} reveals peculiar similarities with specific Nordic countries for adult populations. In particular, the persistent similarity with Danish female log-mortality rates motivates further investigation, as this sub-population is known to have exhibited peculiar mortality behavior due to significant smoking prevalence. Furthermore, the inferred associations between the clustering patterns reconstructed by the proposed model and relevant socio-economic indicators (see Section~\ref{sec_53}), highlight peculiar cohort effects that are worth future analyses.

Our findings in simulations and in the application clarify that clustering at the level of basis coefficients effectively translates into similarities in the observed mortality curves, thereby providing insights into locally-varying demographic structures. Moreover, sensitivity analyses indicate that our results are robust with respect to hyperparameter selection. However, the overall conclusions are clearly related to the selected bases in Figure~\ref{fig:splines}. While this default choice is supported by both demographic practice and successful implementations in mortality modeling, it remains of interest to explore alternative constructions for local clustering of partially overlapping trajectories \citep[e.g.][]{paulon2023bayesian}. Future research includes also applying the proposed model to a broader range of countries, while refining the analysis on the association among the group structures induced by the mortality patterns and key socio-economic indicators. Both perspectives require overcoming challenges related to data availability. In fact, the 14 countries considered  are among the few ones providing mortality rates that date back to our study period. Analyzing more countries would imply constraining the analysis to narrower time windows. Improving inference on the associations with socio-economic indicators would require, instead, including such information directly in the model. A possible direction for addressing this goal is to combine the $t$RPM prior with a reinforcement mechanism favoring the formation of local clusters that are homogeneous also with respect to the associated socio-economic indicators. This could be accomplished by extending the combination among product partition models with covariates \citep{Muller&Quintana&Rosner2011} and $t$RPM priors proposed in  \citet{Page2022} to the case of dynamic covariates. Finally, let us emphasize that although our model is motivated by demographic applications, the constructions in Sections~\ref{section:model_formulation}--\ref{sec_3} have  broader methodological impact, and can be applied whenever interest lies in the  detection of localized overlaps among surfaces associated with different populations. To our knowledge, advancements in these directions are limited.

\section{\bf Acknowledgments}
The authors are grateful to the Editor, the Associate Editor and the referees for the constructive suggestions, which helped in improving the initial version of this article. In addition, the authors would  like to thank Prof.\ Stefano Mazzuco for the insightful discussion on an early version of this manuscript.
\\
\\
{\em Conflicts of interest: } The authors declare that they have no conflicts of interest.

\section{\bf Funding}
This research is supported by the MUR–PRIN 2022 project “CARONTE” (Prot. 2022KBTEBN), funded by the European Union -- Next Generation EU, Mission 4, CUP: J53D23009400001.

\section{\bf Data availability}
The original data analyzed  are available at \url{https://www.mortality.org/}, whereas code can be found at \url{https://github.com/emanuelealiverti/ClustMortSurf}.


\end{document}